\documentclass[prb, aps,twocolumn,superscriptaddress, floatfix]{revtex4-1}

\usepackage{graphicx}
\usepackage{xcolor}
\usepackage{dcolumn}
\usepackage{bm}
\usepackage{amsmath}
\usepackage{amssymb}

\usepackage{ulem}

\begin{document}

\title{Epitaxial stabilization of pulsed laser deposited Sr$_{n+1}$Ir$_n$O$_{3n+1}$ thin films: entangled effect of growth dynamics and strain}

\author{Araceli \surname{Guti\'{e}rrez--Llorente}}
\email[]{araceli.gutierrez@urjc.es}
\affiliation{Universidad Rey Juan Carlos, Escuela Superior de Ciencias Experimentales y Tecnolog\'{i}a, Madrid 28933, Spain}

\author{Luc\'{i}a Iglesias}
\affiliation{Centro Singular de Investigaci\'{o}n en Qu\'{i}mica Biol\'{o}xica e Materiais Moleculares (CiQUS), Departamento de Qu\'{i}mica--F\'{i}sica, Universidade de Santiago de Compostela. 17582 Santiago de Compostela (Spain)}

\author{Benito Rodr\'{i}guez-Gonz\'{a}lez}
\affiliation{Departamento de Qu\'{i}mica F\'{i}sica, Universidad de Vigo, 36310 Vigo, Spain}

\author{Francisco Rivadulla}
\affiliation{Centro Singular de Investigaci\'{o}n en Qu\'{i}mica Biol\'{o}xica e Materiais Moleculares (CiQUS), Departamento de Qu\'{i}mica--F\'{i}sica, Universidade de Santiago de Compostela. 17582 Santiago de Compostela (Spain)}


\begin{abstract}

The subtle balance of electronic correlations, crystal field splitting and spin--orbit coupling in layered Ir$^{4+}$ oxides can give rise to novel electronic and magnetic phases.  Experimental progress in this field relies on the synthesis of epitaxial films of these oxides.  However, the growth of layered iridates with excellent structural quality is a great experimental challenge.  Here we selectively grow high quality single--phase films of Sr$_2$IrO$_4$, Sr$_3$Ir$_2$O$_7$, and SrIrO$_3$ on various substrates from a single Sr$_3$Ir$_2$O$_7$ target by tuning background oxygen pressure and epitaxial strain.  We demonstrate a complex interplay between growth dynamics and strain during thin film deposition.  Such interplay leads to the stabilization of different phases in films grown on different substrates under identical growth conditions, which cannot be explained by a simple kinetic model.  We further investigate the thermoelectric properties of the three phases and propose that weak localization is responsible for the low temperature activated resistivity observed in SrIrO$_3$ under compressive strain.

\end{abstract}

\maketitle 

\section{\label{sec:intro}Introduction}

Transition--metal oxides with partially filled $d$--electron states exhibit exceptionally rich electronic and magnetic phase diagrams.\cite{JBG_MCB,imada:98, tokura:00}  Strong Coulomb interactions in narrow 3$d$--bands can lead to Mott insulators ground states.  Moving down into $5d$ transition metal ions, the spatial extent of the orbitals increases, resulting in a stronger $5d$--O:$2p$ overlap. This widens the electronic bands, thus reducing electronic correlations, whereas the contribution of the crystal--field energy, $\Delta$, increases. Moreover, the higher atomic number results in a significant contribution of the spin--orbit coupling (SOC) energy.  As a result, 5$d$--Iridium (Ir$^{4+}$) oxides constitute a family of materials in which SOC, $\Delta$, and electronic correlations present a comparable magnitude.  Within this context, it has been proposed that SOC splits the threefold degenerate $t_{2g}$ band in Sr$_2$IrO$_4$ into a lower $J_{\mbox{eff}}$=3/2 band, and a half--filled $J_{\mbox{eff}}$=1/2 band. Coulomb interactions or additional structural distortions introduce further splitting in that narrow $J_{\mbox{eff}}$=1/2 band, opening up a charge gap, which can be manipulated by epitaxial strain.\cite{Kim:08,kim:09,PardoPRB,liu:16, okada:13, wang:13,wojek:12,nichols:13,serrao:13,gruenewald:14}

The Ruddlesden-Popper (RP) series of strontium iridates, Sr$_{n+1}$Ir$_n$O$_{3n+1}$, features a localized--to--itinerant crossover from that insulating ground state of Sr$_2$IrO$_4$ (with a two dimensional IrO$_6$ corner--sharing octahedral network characteristic of $n=1$) to a correlated metallic state in the three dimensional perovskite SrIrO$_3$ ($n=\infty$).\cite{moon:08}  This suggests the possibility of fine--tuning the charge gap across Sr$_{n+1}$Ir$_n$O$_{3n+1}$ by growing high quality epitaxially strained thin films.  Theoretically, tensile strain brings about an increase of the Ir--O--Ir bond angle that could favour larger bandwidths and conductivity, at least for $n=1$.\cite{nichols:13} In such case, Serrao et al.\cite{serrao:13} proposed that the charge gap in Sr$_2$IrO$_4$ depends on the ratio $c/a$ of lattice parameters.  But, as $n$ increases the hopping along the $c$--axis becomes more relevant, and thus such a simple analysis might be no longer adequate for other members of the RP series.

High--quality films are required to elucidate these compelling questions, making  indispensable  a  precise  control  of  the  growth  process.  But, as the unit cell of each member of the RP series is a superlattice that consists of ordered sequences of perovskite layers (SrIrO$_3$) sandwiched between two rock salt layers (SrO) along the $c$ crystallographic axis, intergrowth of different members is frequently found.\cite{pallecchi:16,seo:16}  This renders the growth of such artificial layered phases a great experimental challenge, and their growth mechanisms, intriguing.\cite{lee:14, nie:14}

\begin{figure}
 \includegraphics[keepaspectratio=true, width=\linewidth]{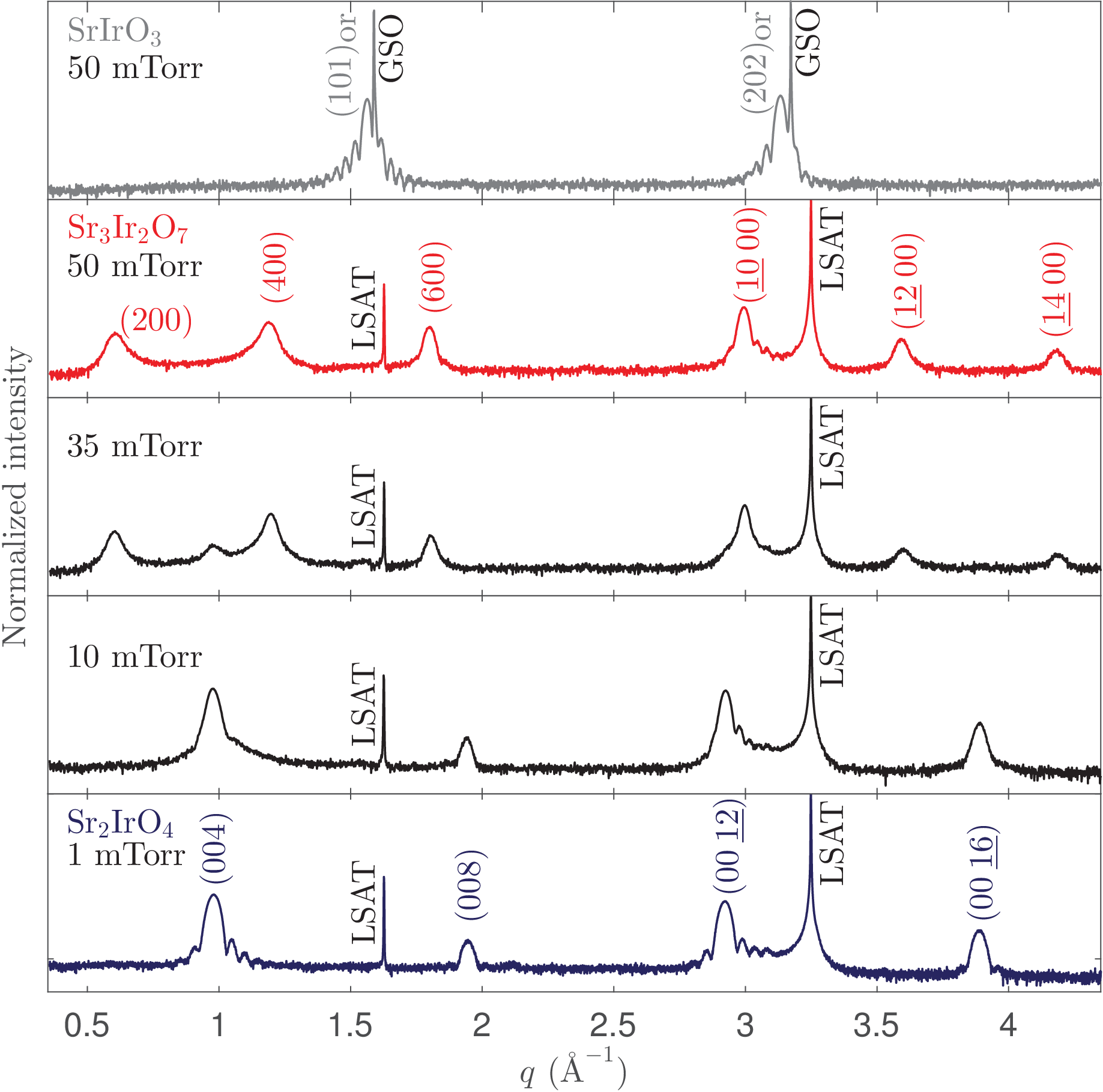} \caption{\label{Fig_01} Epitaxial thin films of $n=1,\;2,\; \infty$ phases of Sr$_{n+1}$Ir$_n$O$_{3n+1}$ series grown from a single Sr$_3$Ir$_2$O$_7$ target. (a) Indexed XRD $\theta$--2$\theta$ scans of films grown on LSAT  over an oxygen pressure range of 1 mTorr to 50 mTorr (from bottom to top), and on GSO at 50 mTorr (upper panel).  Substrate temperature was 800 $^{\circ}$C.  As a rule, increasing the oxygen pressure promotes the stability of phases with larger $n$.  The film grown at 35 mTorr on LSAT shows features that can be attributed to both $n$=1 and $n$=2 RP phases, suggesting a gradual transformation between the two phases.  Noteworthy, Sr$_3$Ir$_2$O$_7$/LSAT and SrIrO$_3$/GSO are both grown on different substrates side by side in the PLD chamber.  (see Fig. S1\cite{supplementary} for structural characterization of Sr$_2$IrO$_4$ and Sr$_3$Ir$_2$O$_7$ phases on STO substrates; Fig. S2\cite{supplementary} for $\theta$--2$\theta$ scans of SrIrO$_3$/GSO films grown at oxygen pressure in the range 35 mTorr to 100 mTorr; and, Fig. S3\cite{supplementary} and Fig. S4\cite{supplementary} that provide evidence of the influence of substrate on the stabilization of RP phases in films grown under identical deposition conditions).  
}
\end{figure}

\begin{figure}
 \includegraphics[keepaspectratio=true, width=\linewidth]{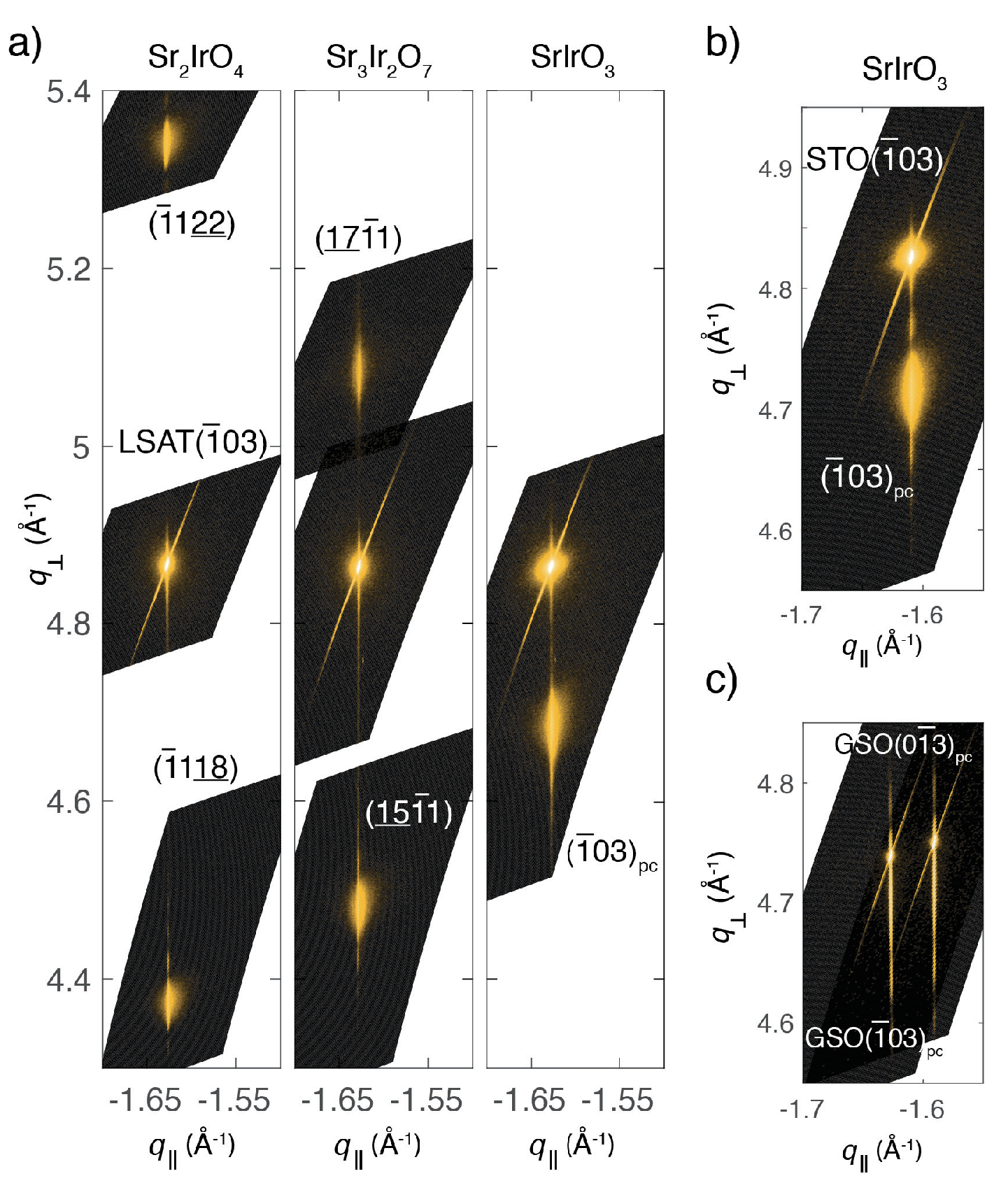} \caption{\label{Fig_02}  a) High--resolution RSMs of Sr$_2$IrO$_4$, Sr$_3$Ir$_2$O$_7$, and SrIrO$_3$ on LSAT.  The $(\bar{1}03)$ reflection from the LSAT substrate is also shown in each map. b, c)  RSMs around the $(\bar{1}03)$ substrate reflection for films grown on STO(001) and GSO(110)$_{\mbox{or}}$, respectively.  The pseudocubic reflections $(\bar{1}03)_{\mbox{pc}}$ and $(0\bar{1}3)_{\mbox{pc}}$ in SrIrO$_3$/GSO are observed at different $(q_{\parallel}, q_{\perp})$ (c).  This is consistent with the deviation of $\beta$ from 90$^{\circ}$ that takes account of the orthorhombic distortion in the perovskite.
}
\end{figure}

Here we report the selective growth of high quality single--phase films of Sr$_2$IrO$_4$, Sr$_3$Ir$_2$O$_7$, and SrIrO$_3$ by pulsed laser deposition (PLD) from a single polycrystalline Sr$_3$Ir$_2$O$_7$ target on substrates that impose different sign and degree of strain.  The growth of RP phases on SrTiO$_3$ substrates from a SrIrO$_3$ or Sr$_2$IrO$_4$ target has been reported by Nishio {\textit{et al}},\cite{nishio:16} and by Liu {\textit{et al}}.\cite{liu:17}, respectively.  In this this work, unlike these previous studies, we explore the impact of the strain imposed by the substrate on the growth of different phases.  We demonstrate a complex interplay between strain and oxygen pressure during growth that can lead to the stabilization of different RP phases in films grown on different substrates under identical deposition conditions of laser fluence, substrate temperature, and background oxygen pressure.  We also discuss the effect of strain, oxygen pressure, and dimensionality on the temperature dependence of thermoelectric power, electrical resistivity, and magnetization of these materials.

\section{\label{sec:results}Results and discussion}

We grew 20 nm thick films of Sr$_{n+1}$Ir$_n$O$_{3n+1}$ with $n$=1, 2 and $\infty$ on (001)SrTiO$_3$ (STO), (001)(LaAlO$_3$)$_{0.3}$(Sr$_2$AlTaO$_6$)$_{0.7}$ (LSAT), and (110)GdScO$_3$ (GSO) substrates by PLD from a single polycrystalline target of Sr$_3$Ir$_2$O$_7$.  The laser fluence was optimized at $\approx$1 J/cm$^2$, and kept constant throughout the work.  (See supplementary material for a detailed experimental description\cite{supplementary}).

\begin{figure*}
 \includegraphics[keepaspectratio=true, scale=0.89]{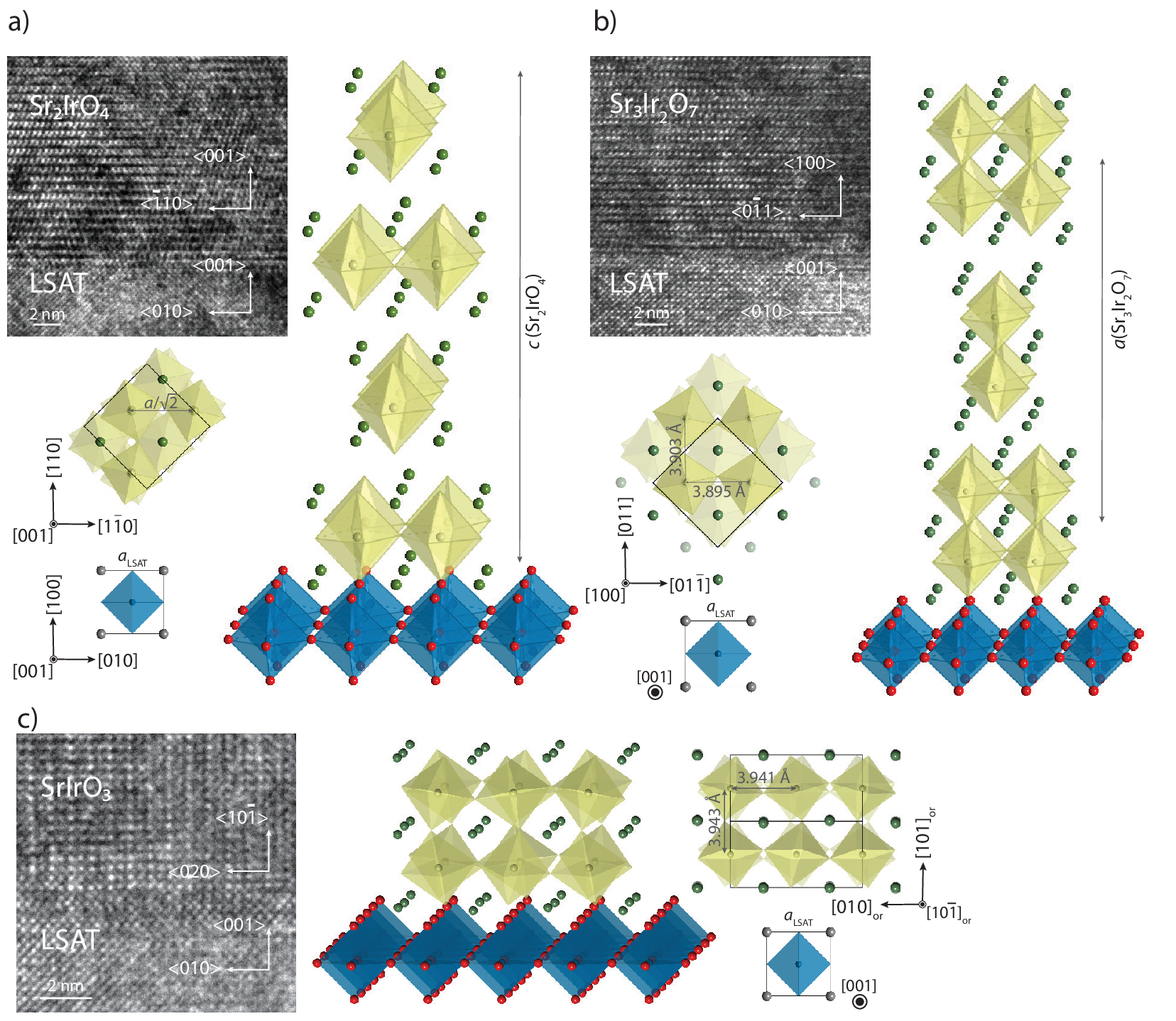}%
\caption{\label{Fig_TEM}High Resolution Transmission Electron Microscopy (HRTEM) images and structural model of the films on LSAT.  Each panel shows the HRTEM image of the interface film/substrate, the 3D structural model, and the in--plane view.  The deduced epitaxial alignments are outlined on each image.  a) Tetragonal\cite{huang:94} Sr$_2$IrO$_4$(001)$\parallel$LSAT(001), where the unit cell of Sr$_2$IrO$_4$ grows 45$^{\circ}$ rotated relative to the LSAT unit cell: $[110]$Sr$_2$IrO$_4$$\parallel$$[100]$LSAT.  This enables lattice matching of $a(\mbox{Sr}_2\mbox{IrO}_4)/\sqrt{2}=$ 3.888 $\mbox{\AA}$ with $a(\mbox{LSAT})$=3.868 $\mbox{\AA}$.  b) Sr$_3$Ir$_2$O$_7$(100)$\parallel$LSAT(001), where Sr$_3$Ir$_2$O$_7$ is depicted as a distorted perovskite (long axis along $a$\cite{hogan:16}).  This refined structure gives rise to an in--plane pseudocubic lattice of roughly 3.90 $\mbox{\AA}$.  The unit cell of Sr$_3$Ir$_2$O$_7$ phase also grows 45$^{\circ}$ rotated relative to the LSAT substrate: $[011]$Sr$_3$Ir$_2$O$_7$$\parallel$$[100]$LSAT.  c) Orthorhombic SrIrO$_3$ ($b$ as long axis\cite{zhao:08,puggioni:16}) shows an epitaxial relationship of $(10\bar{1})_{\mbox{or}}[101]_{\mbox{or}}$SrIrO$_3$$\parallel$$(001)[100]$LSAT.  The tetrahedral rotation patterns expected for each space group in bulk are depicted.  See Supplementary Information for electron diffraction patterns (Fig. S5\cite{supplementary}).}%
\end{figure*}

\begin{figure}
 \includegraphics[keepaspectratio=true, width=\linewidth]{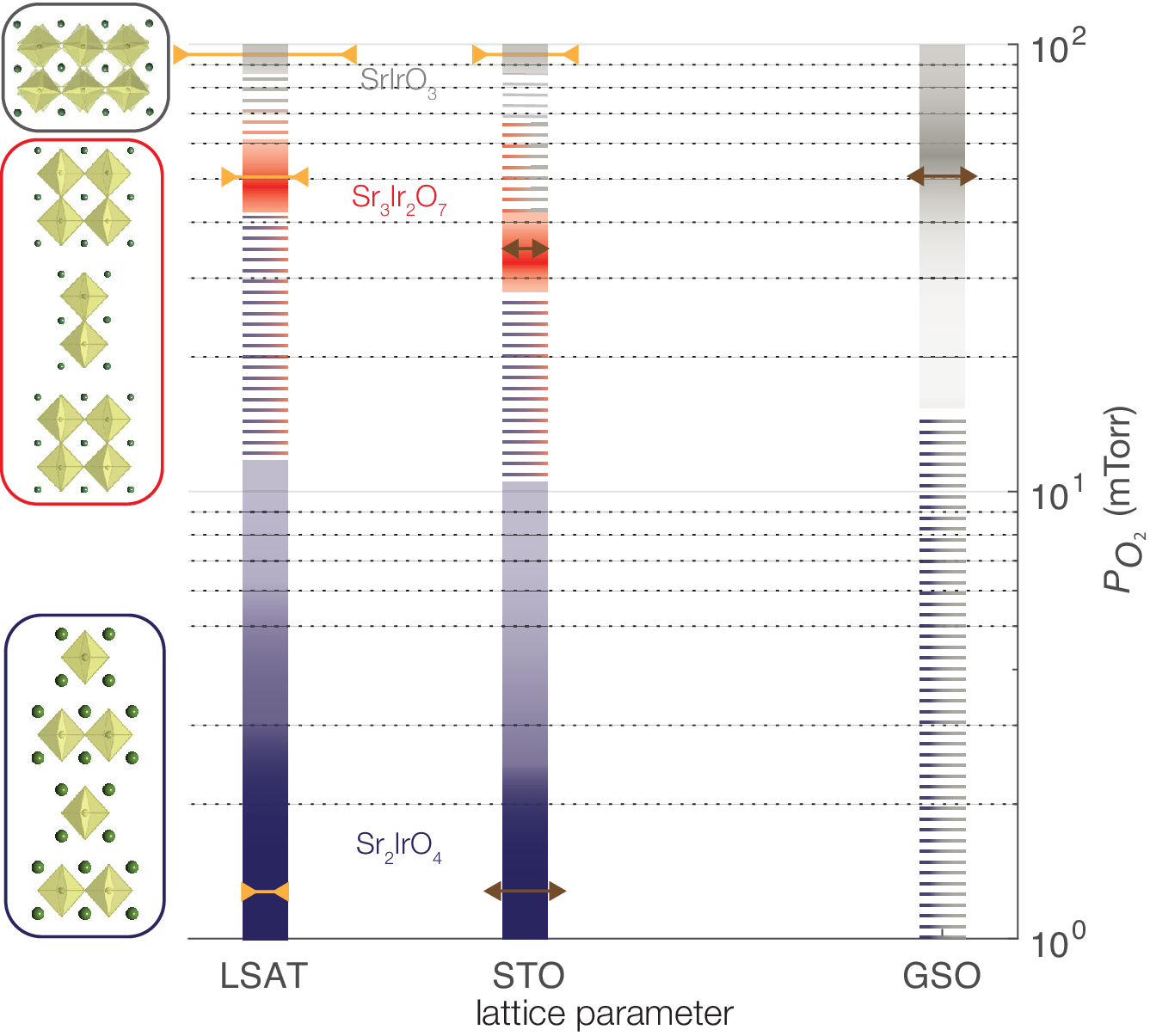} \caption{\label{Fig_growth_conditions} Combined effect of epitaxial strain and background oxygen pressure from a single Sr$_3$Ir$_2$O$_7$ target at a substrate temperature of 800 $^{\circ}$C.  Vertical bars show the experimental oxygen pressure window for the stabilisation of Sr$_2$IrO$_4$ (blue), Sr$_3$Ir$_2$O$_7$ (red) and SrIrO$_3$ (grey) phases on LSAT, STO and GSO substrates.  Films grown over the oxygen pressure range denoted by discontinuous stripes exhibit features of two phases.  We were not able to grow the Sr$_2$IrO$_4$ phase at low oxygen pressure on GSO substrates (see Supplementary Information, Fig. S6\cite{supplementary}).  The pictures on the left illustrate the crystal structures of the three phases.  Compressive (tensile) strain induced by the substrate on each structure is depicted by yellow (brown) arrows whose lengths are proportional to the magnitude of strain.
}
\end{figure}

Off--stoichiometric transfer of material from a multicomponent oxide target to the substrate in pulsed PLD is an undeniable fact.\cite{wicklein:12,seo:16,schraknepper:16}  It is widely accepted that the angular distribution of the species attaining the substrate is modified as a result of the atomic collisions between the atoms of the plume and the molecules of the gas,\cite{packwood:13,sambri:16}  and changes in the oxygen background pressure can lead to a changeover in the growth mode.\cite{groenen:15,tselev:16}  Furthermore, epitaxial strain has a major impact on growth dynamics as it modifies surface diffusion and mobility of adatoms, although the character of that shift is strongly material dependent, and has been little explored in oxides.\cite{brune:95,schroeder:97,xu:14,pandya:16,tselev:16}

Fig. \ref{Fig_01} shows x--ray diffraction (XRD) $\theta$--2$\theta$ scans of films grown on LSAT at an oxygen pressure of 1 mTorr, 10 mTorr, 35 mTorr and 50 mTorr, and on GSO at 50 mTorr, at a substrate temperature of 800 $^{\circ}$C.  The patterns of films grown at 1 mTorr and 50 mTorr on LSAT, and at 50 mTorr on GSO show all the peaks of Sr$_2$IrO$_4$, Sr$_3$Ir$_2$O$_7$, and SrIrO$_3$, respectively, oriented with the long axis along the substrate normal.  There is no hint of impurity phases in these patterns which suggests that the sequence of perovskite and rocksalt layers are correctly ordered in the periodic structures along the out--of--plane direction (see discussion of Transmission Electron Microscopy data below).  Strong Laue thickness fringes surrounding the main Bragg peak give evidence of an excellent structural quality of these films.  Rocking curve measurements for the (00$\underline{12}$)Sr$_2$IrO$_4$ peak show a FWHM of 0.02$^{\circ}$$\pm$0.003$^{\circ}$ and 0.06$^{\circ}$ $\pm$0.005$^{\circ}$ for the films grown on LSAT substrates under 1 mTorr and 10 mTorr, respectively.  The FWHM of the (002)$_{\mbox{or}}$SrIrO$_3$ peak of the film grown on GSO under 50 mTorr is 0.05$^{\circ}$$\pm$0.003$^{\circ}$.  The diffraction peaks of Sr$_2$IrO$_4$ can be indexed to the tetragonal unit cell: $a=5.499\: \mbox{\AA}$ and $c=25.784 \: \mbox{\AA}$, space group I4$_1$/$acd$.\cite{huang:94}  The peaks observed in Sr$_3$Ir$_2$O$_7$  are compatible with a subtly distorted perovskite described by a monoclinic space group ($C2/c$)\cite{hogan:16} with parameters 20.935 $\mbox{\AA}$, 5.5185 $\mbox{\AA}$, and 5.5099 $\mbox{\AA}$, where the long cell axis ($a$ in the standard setting of space group $C2/c$) is parallel to the surface normal. In this structure, a single oblique angle $\beta$=90.045${^\circ}$ produces a minute deviation from a quadratic lattice\cite{hogan:16}, which is below our X--ray experimental resolution.  SrIrO$_3$ has been indexed to an orthorhombic perovskite cell with $Pnma$ symmetry and lattice parameters $a=5.5909\: \mbox{\AA}$, $b=7.8821\: \mbox{\AA}$, and $c=5.5617\: \mbox{\AA}$\cite{zhao:08,puggioni:16} (pseudo--cubic unit cell of roughly $3.94\: \mbox{\AA}$). 

X--ray reciprocal space maps (RSMs) of the three phases on different substrates in Fig. \ref{Fig_02} show that the films are fully strained to the in--plane lattice parameter of the substrate.  As expected, the out--of--plane lattice parameter of Sr$_2$IrO$_4$ expands (shrinks) relative to that of bulk under in--plane compressive (tensile) strain induced by LSAT (STO) substrate.  It decreases from 25.86$\pm$0.05 $\mbox{\AA}$ on LSAT  to 25.72$\pm$0.03 $\mbox{\AA}$ on STO (tensile strain of $\approx$ $+0.64\%$).  Likewise, in--plane tensile (compressive) strain also results in a contraction (expansion) along the out-of-plane direction in Sr$_3$Ir$_2$O$_7$.  It shrinks along the c--axis from 21.03$\pm$0.03 $\mbox{\AA}$ under compressive strain on LSAT to 20.81$\pm$0.04 $\mbox{\AA}$  under tensile strain on STO.  These values are in good agreement with values for the bulk of $n$=1 and $n$=2 RP phases.\cite{huang:94,subramanian:94}

Further insight into the microstructure and composition of the films is provided by Transmission Electron Microscopy (TEM) on cross--section lamellas (Fig. \ref{Fig_TEM}). In agreement with the RSM results, we find that the films are fully strained to the substrate.  The analysis of the electron diffraction patterns (Fig. S5\cite{supplementary}) reveals in--plane orientation relationships for Sr$_2$IrO$_4$ and Sr$_3$Ir$_2$O$_7$ films where the unit cell of the iridate grows 45$^{\circ}$ rotated relative to the LSAT unit cell (Fig. \ref{Fig_TEM}(a, b)).  Orthorhombic SrIrO$_3$ ($b$ as long axis) shows an epitaxial relationship of $(10\bar{1})_{\mbox{or}}[101]_{\mbox{or}}$SrIrO$_3$$\parallel$$(001)[100]$LSAT (Fig. \ref{Fig_TEM}(c)).  Semiquantitative Energy Dispersive Spectroscopy (EDS) analyses of the films confirmed a close to stoichiometric Sr/Ir ratio of 1.91, 1.74, and 1.12 for films grown under oxygen pressure of 1 mTorr, 50 mTorr and 100 mTorr, respectively.

Fig. \ref{Fig_growth_conditions} summarizes the entanglement of epitaxial strain and background oxygen pressure in the stabilization of Sr$_{n+1}$Ir$_n$O$_{3n+1}$ phases.  For example,  Sr$_2$IrO$_4$ grows with excellent structural quality on LSAT (small lattice mismatch $\approx$ $-0.26$\%) at oxygen pressure about 1 mTorr, as we have stated above.  However, this phase does not stabilize under a strain of $\approx$ $+2$\% (GSO substrate) over the range of oxygen pressures studied in this work (see Fig. S6\cite{supplementary}).  Furthermore, we obtain single--phase Sr$_3$Ir$_2$O$_7$ films at 50 mTorr on LSAT (lattice mismatch of $\approx$ $-0.75$\%); and, at 35 mTorr on STO (lattice mismatch $\approx$ $+0.15$\%, Fig. S1\cite{supplementary}).  Moreover, SrIrO$_3$ grows with high structural quality tensile--strained ($\approx$ $+0.51$\%) on GSO already above $\approx$ 40 mTorr (see also Supplementary Information, Fig. S2\cite{supplementary}).

A picture emerges from these results: epitaxial stress plays a fundamental role in the stabilization of RP phases; and, low (high) oxygen partial pressure favours the formation of phases with high (low) Sr/Ir ratio.  PLD plume dynamics can be invoked to give a first interpretation of these results.  In a simple kinetic model, the plume species are expected to diffuse while interacting with the background gas until reaching the substrate.  Given the large difference in size and mass between Ir and Sr, it is expected that background oxygen pressure will significantly affect propagation velocity and angular distribution of Ir and Sr species which, in turn, will impact on the Sr/Ir cation ratio of the films.

At the working laser fluence of $\approx$1 J/cm$^2$, we observe that Sr is preferentially ablated from the target, as proved by EDS analysis carried out on the target.  The original stoichiometric Sr/Ir ratio is recovered after careful polishing of the surface of the target.  We grew the films after a long preablation of the target surface to ensure that a steady state had been reached.  Such a preferential ablation of Sr has previously been reported for a SrIrO$_3$ target at laser fluences varying from 0.4 to 2.0 J/cm$^2$.\cite{groenendijk:16}

At low background oxygen pressure, it is predicted that lighter Sr species outnumber heavier Ir species at the plume front.\cite{sambri:16}  This effect helps overcoming the Ir enrichment of the target surface owing to Sr preferential ablation.  In fact, results of complementary growths hint that preferential ablation of Sr is not a key condition for the stabilization of the Sr$_2$IrO$_4$ phase at low oxygen pressure (see Fig. S7\cite{supplementary}, and discussion in Supplementary Information).  As a result, films with Sr/Ir ratio higher than that of the Sr$_3$Ir$_2$O$_7$ polycrystalline target are grown at oxygen pressure around 1 mTorr on LSAT and STO substrates.  As the oxygen pressure inside the chamber increases, the propagation behaviour of the species changes.\cite{groenen:15}  In this case, lighter and larger species, such as Sr, are preferentially scattered during their flight towards the substrate. Therefore, as the number of scattering events increases at high oxygen pressure, there is an enrichment in Ir along the direction normal to the substrate that leads to films with Sr/Ir ratio lower than that of the Sr$_3$Ir$_2$O$_7$ target.

\begin{figure}
 \includegraphics[keepaspectratio=true, width=\linewidth]{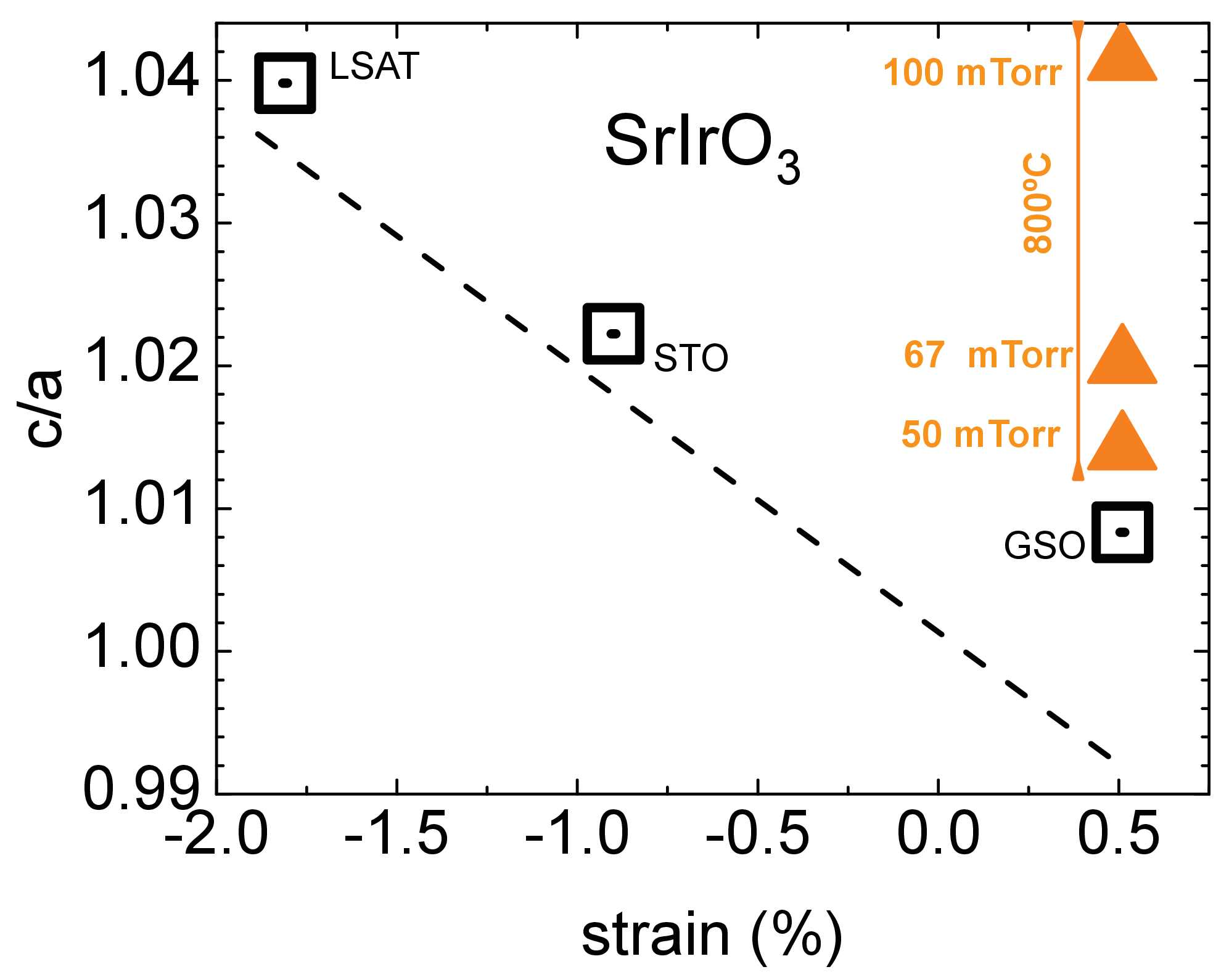}%
\caption{\label{Fig_parameters_SrIrO3}Variation of the $c/a$ ratio of pseudocubic lattice parameters with strain (open squares) and background oxygen pressure (solid triangles) of SrIrO$_3$ films.  The open squares correspond to samples synthesized at 67 mTorr and 700$^{\circ}$C in the same batch. The triangles correspond to samples deposited on GSO  at 800$^{\circ}$C, under different oxygen pressures (this data are extracted from XRD $\theta$--2$\theta$ scans shown in Fig. S2 and Fig. S8\cite{supplementary}).  The dotted line corresponds to $c/a$ calculated for a Poisson's ratio of $\nu$=0.3.}
\end{figure}

Fig. \ref{Fig_parameters_SrIrO3} shows the evolution of unit cell parameters of SrIrO$_3$ as function of strain for samples synthesized under different temperatures and oxygen pressures.  The values shown in Fig. \ref{Fig_parameters_SrIrO3} are in accordance with literature.\cite{nie:15,gruenewald:14,groenendijk:16}  Under the optimized growth conditions, the tetragonal distortion, $c/a$, decreases with tensile strain, although at a lower rate than that expected assuming a Poisson's ratio of $\nu=0.3$, a value common to other oxide--perovskites.\cite{greaves:11, huang_Poisson:16}  We find that the $c/a$ ratio increases with increasing temperature (oxygen pressure) while keeping the oxygen pressure (temperature) constant.  For instance, the film grown at the highest pressure (100 mTorr) exhibits the highest $c/a$ ratio.  Thus, we exclude the presence of oxygen vacancies as the cause of the deformation of the unit cell of SrIrO$_3$ depicted in Fig. \ref{Fig_parameters_SrIrO3}.\cite{iglesias:17, aschauer:13}  Instead the behaviour of $c/a$ has to be caused by an increase of cation vacancies with increasing temperature and/or oxygen pressure.  The preferential scattering of Sr in the plume at high oxygen pressure predicted by the kinetic model described above would result in SrIrO$_3$ films with an increased concentration of Sr vacancies, giving rise to the cell expansion observed in Fig. \ref{Fig_parameters_SrIrO3} for the films grown under tensile strain.

In adittion, transport properties of epitaxially grown SrIrO$_3$ films are significantly affected by non--stoichiometric Sr/Ir ratio.  Indeed, the SrIrO$_3$ film grown on GSO at 800$^{\circ}$C and 100 mTorr of oxygen pressure exhibits semiconducting--like behaviour (Fig. S9\cite{supplementary}) which can be associated with the increase of Sr vacancies postulated above.  A film of SrIrO$_3$ grown from a target with higher Sr/Ir ratio under identical conditions of substrate temperature and oxygen pressure exhibits metallic behaviour (Fig. S10.\cite{supplementary}), supporting the picture of Anderson--like localization by vacancy scattering.

On the other hand, such a simple model of PLD plume dynamics overlooks the influence of epitaxial strain on diffusion of species at the substrate surface.  Actually, we observe that epitaxial strain can promote the stabilization of different phases in films grown on different substrates under identical growth conditions of laser fluence, substrate temperature, and oxygen pressure (Fig. \ref{Fig_01} and Fig. S4\cite{supplementary}).   For instance, it has been reported that tensile epitaxial strain brings about an energy barrier for adatom diffusion during the growth of complex oxides at low oxygen pressures by PLD.\cite{tselev:16}  We hypothesize this decrease of surface diffusion and mobility of adatoms hinders the growth of Sr$_2$IrO$_4$ and Sr$_3$Ir$_2$O$_7$ phases on GSO substrates.  It is also worth bearing in mind that SrIrO$_3$ has lower lattice mismatch with GSO, $\approx$ $+0.51$\%, than Sr$_2$IrO$_4$ ($\approx$ $+2$\%) or Sr$_3$Ir$_2$O$_7$ ($\approx$ $+1.6$\%), providing the driving force for the stabilization of SrIrO$_3$ if growth conditions of substrate temperature or oxygen pressure make it possible.

\begin{figure}
 \includegraphics[keepaspectratio=true, width=\linewidth]{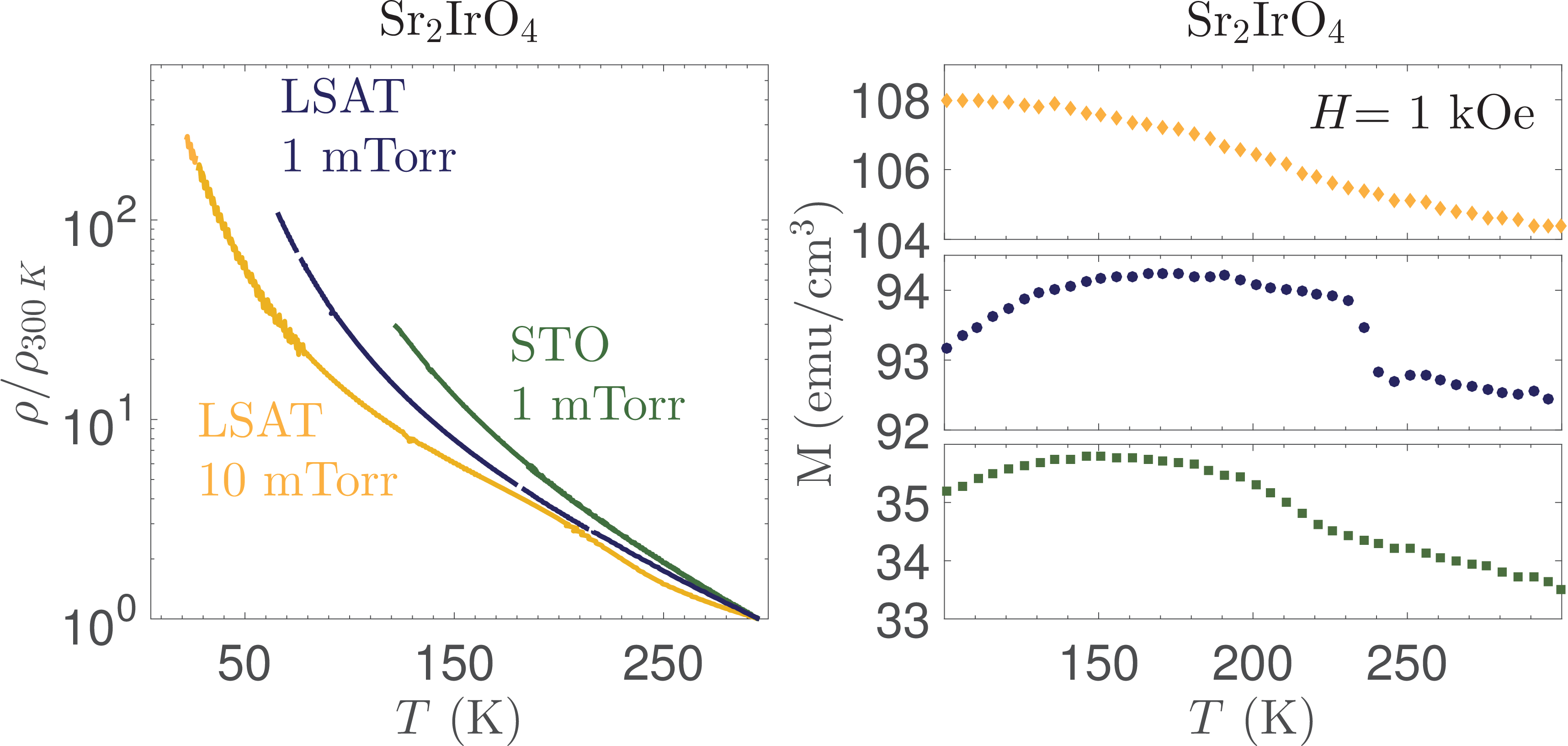}%
\caption{\label{Fig_magnetization_b}Left panel: temperature dependence of electrical resistivity, $\rho(T)$, normalized to room temperature of Sr$_2$IrO$_4$ films grown at 800$^{\circ}$C on LSAT at 1 mTorr (blue), LSAT at 10 mTorr (yellow), and STO at 1 mTorr (green).  Right panel: temperature dependence of magnetization, $M(T)$, at $H$= 1 kOe of the same films.}
\end{figure}

Therefore, both kinetic and thermodynamic aspects are relevant to explain the stabilization of the different phases reported in this work (see supplementary material\cite{supplementary} for a  description of the impact of volatile IrO$_2$).  Additional experiments to probe the angular distribution of Sr and Ir species in the plume, and to further understand the role played by oxygen on the plume propagation dynamics and on the incorporation of volatile species into the film would be highly interesting.

 Fig. \ref{Fig_magnetization_b} summarizes the effect of background oxygen pressure, and strain on magnetic and transport properties of Sr$_2$IrO$_4$.  This phase remains semiconducting irrespective of oxygen pressure and strain, not showing any sensitivity to the magnetic transition, as expected.\cite{chikara:10, shimura:95}  In contrast, as the canting of Ir magnetic moments in Sr$_2$IrO$_4$ follow the IrO$_6$ octahedral rotations,\cite{boseggia:13,ge:11,ye:13} epitaxial strain is expected to have a strong influence on its magnetic properties.  We observe a sharp magnetic transition around $T\approx240$ K (bulk value\cite{kim:09}) for the film grown compressively strained on LSAT at the lowest oxygen pressure, 1 mTorr.  This transition is flattened in the film grown on LSAT at 10 mTorr, resulting from a decreased structural quality of the films grown at higher pressures.  Indeed, rocking curves around the (00$\underline{12}$)Sr$_2$IrO$_4$ peak exhibit a FWHM of 0.02$^{\circ}$$\pm$0.003$^{\circ}$ or 0.06$^{\circ}$ $\pm$0.005$^{\circ}$ for films grown under 1 mTorr or 10 mTorr, respectively, while no significant difference was found in the out--of--plane lattice parameter between both films (25.86$\pm$0.05 $\mbox{\AA}$ and 25.85$\pm$0.06 $\mbox{\AA}$, respectively).  We also find that the ferromagnetic component, which stems from the canted antiferromagnetic order, is lower in the film grown on STO (tensile strain) than in those grown on LSAT (compressive strain).  This is in accordance with previous studies on 200 nm--thick Sr$_2$IrO$_4$ films where higher tensile strain was reported to lead to lower octahedral rotation, resulting in a weaker ferromagnetic component.\cite{miao:14}

\begin{figure}
 \includegraphics[keepaspectratio=true, width=\linewidth]{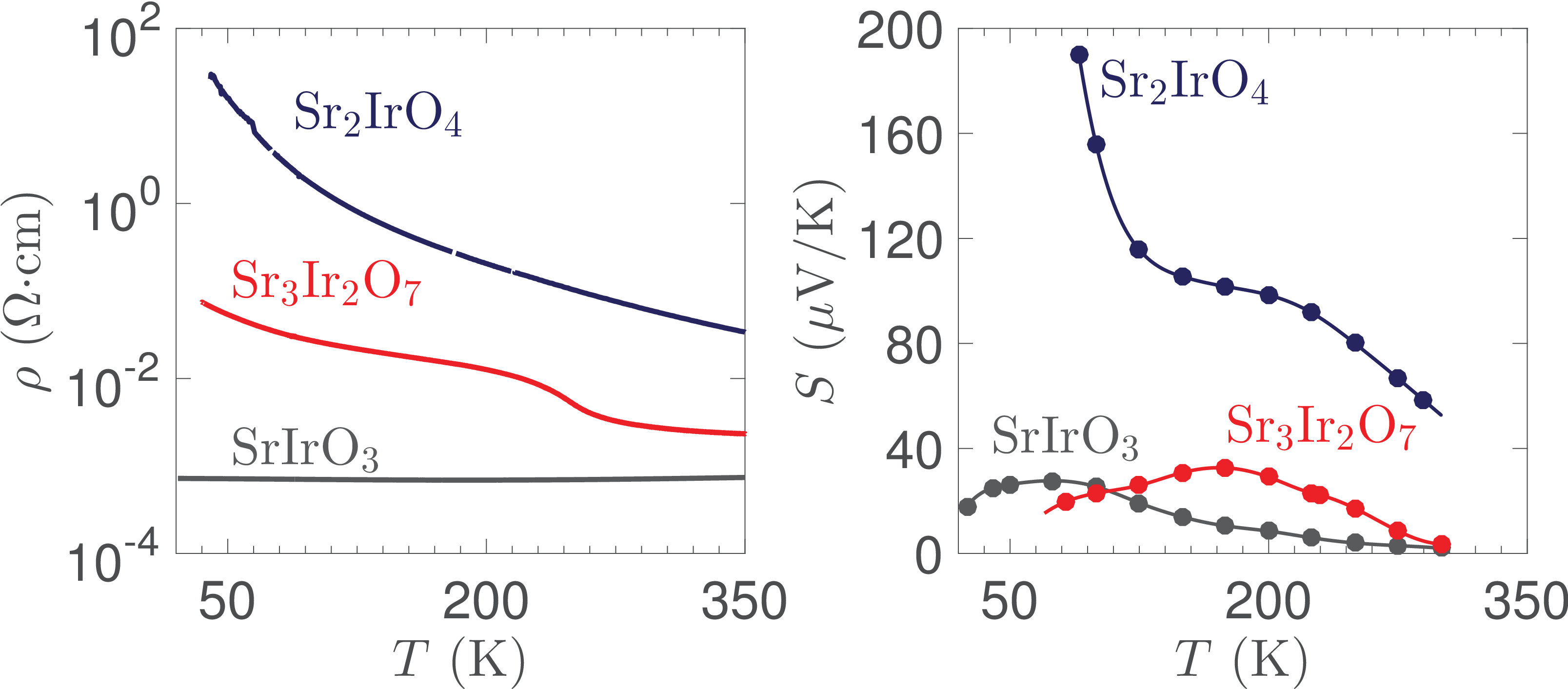}%
\caption{\label{Fig_transport_01}  Temperature dependence of the electrical resistivity, $\rho(T)$, and Seebeck coefficient, $S(T)$, of Sr$_2$IrO$_4$, Sr$_3$Ir$_2$O$_7$, and SrIrO$_3$ films grown on LSAT(001).  The lines in the Seebeck figure are a guide to the eye.}%
\end{figure}

The temperature dependence of resistivity, $\rho(T)$, and Seebeck coefficient, $S(T)$, of Sr$_2$IrO$_4$, Sr$_3$Ir$_2$O$_7$, and SrIrO$_3$ on LSAT are shown in Fig. \ref{Fig_transport_01}.  Epitaxial compression on LSAT increases with increasing dimensionality of the material, $n$.  Electrical transport of these phases is consistent with the widely accepted bandwidth--driven insulator--to--metal transition previously reported on the Sr$_{n+1}$Ir$_n$O$_{3n+1}$ series as a result of increasing dimensionality, $n$: Sr$_2$IrO$_4$/LSAT(001) shows semiconducting behavior and high resistivity; Sr$_3$Ir$_2$O$_7$/LSAT(001) exhibits a semiconducting--like behavior over the whole range of temperature with a characteristic feature that reflects the antiferromagnetic transition expected at $T=285$ K in bulk;\cite{cao:02}  and,  SrIrO$_3$/LSAT(001) presents a very low resistivity and metallic behavior, with a slight upturn at low temperatures.

The Seebeck coefficient of the films on LSAT is positive in the whole range of temperatures measured, Fig. \ref{Fig_transport_01}.  The magnitude of $S(T)$ in Sr$_2$IrO$_4$ is in accordance with previous reports on polycrystalline \cite{kini:06,pallecchi:16} and  single--crystal samples.\cite{chikara:10,shimura:95}  We observed a clear plateau below $\approx$200 K for the epitaxial films that could be linked to the canted antiferromagnetic structure reported in bulk,\cite{kim:09,boseggia:13} although no anomaly of the Seebeck coefficient at the magnetic transition temperature has been observed in single crystals\cite{chikara:10,shimura:95} or polycrystalline\cite{kini:06,pallecchi:16} Sr$_2$IrO$_4$.  We are not aware of any other measurements of the thermoelectrical properties of epitaxial films of Sr$_2$IrO$_4$ or Sr$_3$Ir$_2$O$_7$.

\begin{figure}
 \includegraphics[keepaspectratio=true, width=\linewidth]{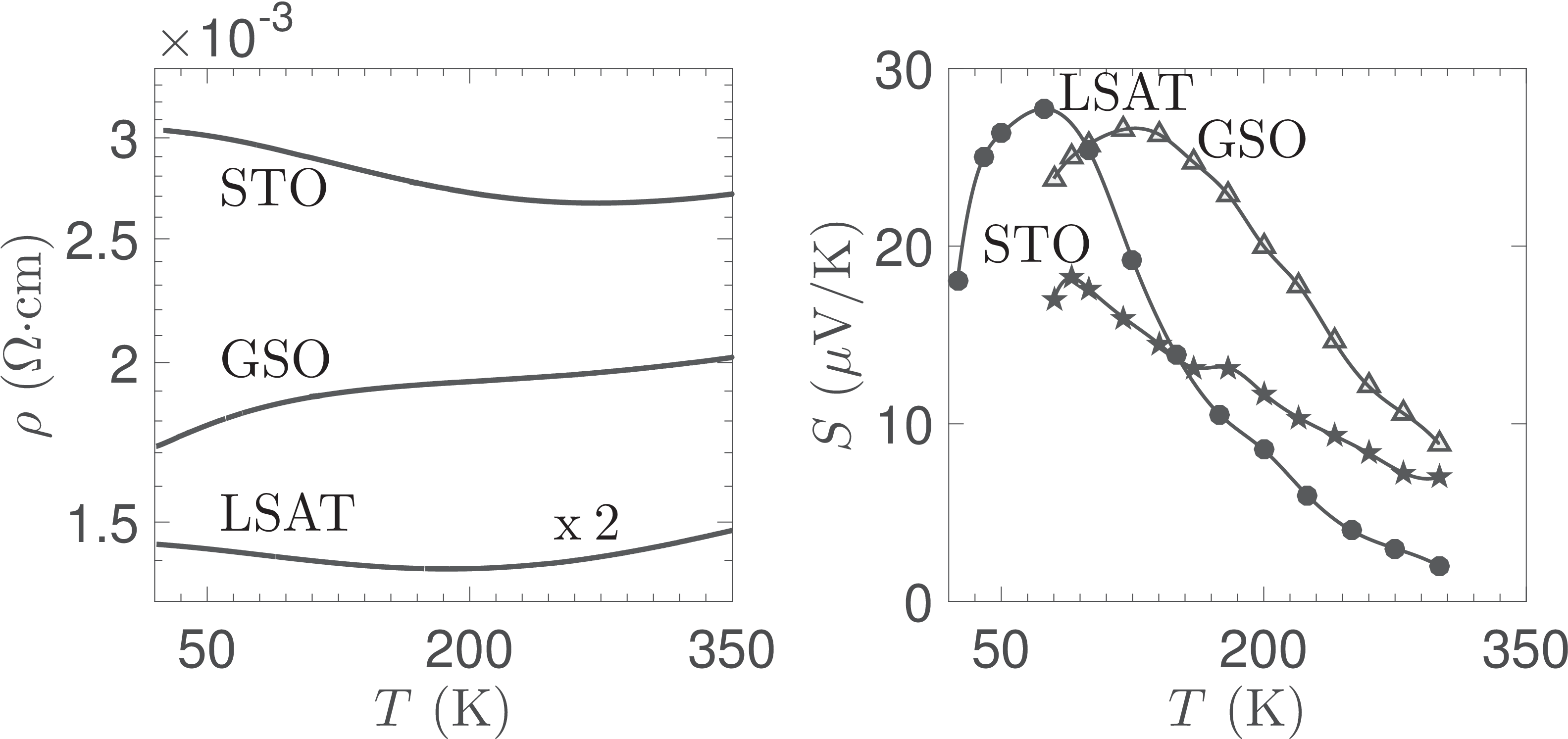}%
\caption{\label{Fig_transport_02}  Temperature dependence of the electrical resistivity, $\rho(T)$, and Seebeck coefficient, $S(T)$, of SrIrO$_3$ films grown on GSO ($\bigtriangleup$), STO ($*$) and LSAT ($\circ$).  These films were grown in the same batch (67 mTorr, 700$^{\circ}$C). The resistivity of the film grown on LSAT has a scale factor of 2.  The lines in the Seebeck figure are a guide to the eye. (See Fig. S8\cite{supplementary} for XRD measurements of the films).}%
\end{figure}

The effect of epitaxial strain on the thermoelectric properties of SrIrO$_3$ is shown in Fig. \ref{Fig_transport_02}.  These films were prepared in the same batch.  Thus, the dissimilarity in their electronic properties must stem from different substrate induced strain: $-1.8\%$, $-0.9\%$ and $+0.5\%$ for LSAT, STO and GSO, respectively.  Moreover, electrical transport measurements were carried out immediately after growth for three days in a row hence, degradation effects reported to occur in the SrIrO$_3$ films\cite{groenendijk:16} are expected to be negligible.  The temperature dependence of $\rho(T)$ shows metallic behavior at high temperature, with a crossover towards a thermally activated state defined by ($d \rho(T)/dT)<0$ at low temperature in the samples under compressive strain.  The magnitude of the $\rho(T)$ and the value of the crossover temperature do not follow any clear dependence with the $c/a$ ratio. Therefore, they cannot be directly related to a change in bandwidth with strain.

This is in contrast to previous studies about the strain effect on electrical transport in SrIrO$_3$ films.\cite{gruenewald:14,biswas:14}  These studies suggested a bandwidth controlled via strain model, according to which the Ir--O--Ir angle decreases (increases) by compressive (tensile) strain (while Ir--O length is not modified), thus shrinking (expanding) the bandwidth.  Such scenario echoes the behavior observed in Sr$_2$IrO$_4$,\cite{serrao:13} but essential differences between Sr$_2$IrO$_4$ and SrIrO$_3$ have been pointed out.  In particular, SrIrO$_3$ exhibits out--of--plane octahedral rotations along the [110] pseudocubic direction, but no such [110] rotations are found experimentally in Sr$_2$IrO$_4$.\cite{nie:15} Furthermore, a reported narrower bandwidth in SrIrO$_3$ than in Sr$_2$IrO$_4$ casts doubt on the conventional picture of increased bandwidth with increasing dimensionality, $n$, in the RP series of iridates.\cite{nie:15}  On the other hand, the Seebeck coefficients are very similar for all SrIrO$_3$ films, with only a slight dependence on epitaxial strain, Fig. \ref{Fig_transport_02}.  As thermoelectric voltage is measured in open circuit conditions, no electrical current flows through the sample, and consequently $S(T)$ is not as sensitive to grain boundaries and point--defect scattering as electrical resistivity.  Indeed, $S(T)$ rather depends on the intrinsic electronic structure of the conductor.  Therefore, we suggest that localization induced by disorder is responsible for the temperature dependence of the resistivity observed in these films. This is in agreement with previous observations of a persistent Drude--like peak in the optical conductivity of SrIrO$_3$ films under comparable strain.\cite{gruenewald:14}

\section{\label{sec:conclusions}Conclusions}

Our results reveal an intricate coupling between epitaxial strain and oxygen pressure during pulsed laser deposition of Sr$_{n+1}$Ir$_n$O$_{3n+1}$ films.  Such coupling suggests the possibility of growing artificial superlattices of iridates with tailored electronic and magnetic properties by varying the background pressure during deposition.\cite{matsuno:15,ohuchi:18}  In addition, future works could address atomic--scale effects of strain on the surface diffusion of species during growth, and the influence of the oxidation state of the arriving species to promote the stabilization of RP phases with different cation stoichiometry.\cite{lee:14,nie:14,tselev:16, tselev:15}  This study would require {\textit{in--situ}} characterization to monitor the film growth, and would provide fundamental insights into the growth process that may lead to the stabilization of new phases of layered materials.  We have also shown that weak localization effects owing to accommodation to compressive epitaxial strain dominate the conductivity of the metallic SrIrO$_3$ films, beyond the simple model of bandwidth controlled via strain.

\section*{\label{sec:suppl_mat}Supplementary Material}

See supplementary material for a detailed experimental description, additional figures, and a comment on the impact of volatile IrO$_2$ .

\begin{acknowledgments}
This work has received financial support from Ministerio de Econom\'{i}a y Competitividad (Spain) under project No. MAT2016--80762--R, Xunta de Galicia (Centro singular de investigaci\'{o}n de Galicia accreditation 2016--2019, ED431G/09) and the European Union (European Regional Development Fund--ERDF).

A.G.L. acknowledges financial support from Universidad de Santiago de Compostela. L.I. also acknowledges Mineco--Spain for support under a FPI grant.

\end{acknowledgments}


\begin{thebibliography}{54}%
\makeatletter
\providecommand \@ifxundefined [1]{%
 \@ifx{#1\undefined}
}%
\providecommand \@ifnum [1]{%
 \ifnum #1\expandafter \@firstoftwo
 \else \expandafter \@secondoftwo
 \fi
}%
\providecommand \@ifx [1]{%
 \ifx #1\expandafter \@firstoftwo
 \else \expandafter \@secondoftwo
 \fi
}%
\providecommand \natexlab [1]{#1}%
\providecommand \enquote  [1]{``#1''}%
\providecommand \bibnamefont  [1]{#1}%
\providecommand \bibfnamefont [1]{#1}%
\providecommand \citenamefont [1]{#1}%
\providecommand \href@noop [0]{\@secondoftwo}%
\providecommand \href [0]{\begingroup \@sanitize@url \@href}%
\providecommand \@href[1]{\@@startlink{#1}\@@href}%
\providecommand \@@href[1]{\endgroup#1\@@endlink}%
\providecommand \@sanitize@url [0]{\catcode `\\12\catcode `\$12\catcode
  `\&12\catcode `\#12\catcode `\^12\catcode `\_12\catcode `\%12\relax}%
\providecommand \@@startlink[1]{}%
\providecommand \@@endlink[0]{}%
\providecommand \url  [0]{\begingroup\@sanitize@url \@url }%
\providecommand \@url [1]{\endgroup\@href {#1}{\urlprefix }}%
\providecommand \urlprefix  [0]{URL }%
\providecommand \Eprint [0]{\href }%
\providecommand \doibase [0]{http://dx.doi.org/}%
\providecommand \selectlanguage [0]{\@gobble}%
\providecommand \bibinfo  [0]{\@secondoftwo}%
\providecommand \bibfield  [0]{\@secondoftwo}%
\providecommand \translation [1]{[#1]}%
\providecommand \BibitemOpen [0]{}%
\providecommand \bibitemStop [0]{}%
\providecommand \bibitemNoStop [0]{.\EOS\space}%
\providecommand \EOS [0]{\spacefactor3000\relax}%
\providecommand \BibitemShut  [1]{\csname bibitem#1\endcsname}%
\let\auto@bib@innerbib\@empty
\bibitem [{\citenamefont {Goodenough}(1963)}]{JBG_MCB}%
  \BibitemOpen
  \bibfield  {author} {\bibinfo {author} {\bibfnamefont {J.~B.}\ \bibnamefont
  {Goodenough}},\ }\href@noop {} {\emph {\bibinfo {title} {Magnetism and the
  Chemical Bond}}}\ (\bibinfo  {publisher} {John Willey and Sons},\ \bibinfo
  {address} {New York},\ \bibinfo {year} {1963})\BibitemShut {NoStop}%
\bibitem [{\citenamefont {Imada}\ \emph {et~al.}(1998)\citenamefont {Imada},
  \citenamefont {Fujimori},\ and\ \citenamefont {Tokura}}]{imada:98}%
  \BibitemOpen
  \bibfield  {author} {\bibinfo {author} {\bibfnamefont {M.}~\bibnamefont
  {Imada}}, \bibinfo {author} {\bibfnamefont {A.}~\bibnamefont {Fujimori}}, \
  and\ \bibinfo {author} {\bibfnamefont {Y.}~\bibnamefont {Tokura}},\ }\href
  {\doibase 10.1103/RevModPhys.70.1039} {\bibfield  {journal} {\bibinfo
  {journal} {Reviews of Modern Physics}\ }\textbf {\bibinfo {volume} {70}}
  (\bibinfo {year} {1998}),\ 10.1103/RevModPhys.70.1039}\BibitemShut {NoStop}%
\bibitem [{\citenamefont {Tokura}\ and\ \citenamefont
  {Nagaosa}(2000)}]{tokura:00}%
  \BibitemOpen
  \bibfield  {author} {\bibinfo {author} {\bibfnamefont {Y.}~\bibnamefont
  {Tokura}}\ and\ \bibinfo {author} {\bibfnamefont {N.}~\bibnamefont
  {Nagaosa}},\ }\href {\doibase 10.1126/science.288.5465.462} {\bibfield
  {journal} {\bibinfo  {journal} {Science}\ }\textbf {\bibinfo {volume} {288}}
  (\bibinfo {year} {2000}),\ 10.1126/science.288.5465.462}\BibitemShut
  {NoStop}%
\bibitem [{\citenamefont {Kim}\ \emph {et~al.}(2008)\citenamefont {Kim},
  \citenamefont {Jin}, \citenamefont {Moon}, \citenamefont {Kim}, \citenamefont
  {Park}, \citenamefont {Leem}, \citenamefont {Yu}, \citenamefont {Noh},
  \citenamefont {Kim}, \citenamefont {Oh}, \citenamefont {Park}, \citenamefont
  {Durairaj}, \citenamefont {Cao},\ and\ \citenamefont {Rotenberg}}]{Kim:08}%
  \BibitemOpen
  \bibfield  {author} {\bibinfo {author} {\bibfnamefont {B.~J.}\ \bibnamefont
  {Kim}}, \bibinfo {author} {\bibfnamefont {H.}~\bibnamefont {Jin}}, \bibinfo
  {author} {\bibfnamefont {S.~J.}\ \bibnamefont {Moon}}, \bibinfo {author}
  {\bibfnamefont {J.~Y.}\ \bibnamefont {Kim}}, \bibinfo {author} {\bibfnamefont
  {B.~G.}\ \bibnamefont {Park}}, \bibinfo {author} {\bibfnamefont {C.~S.}\
  \bibnamefont {Leem}}, \bibinfo {author} {\bibfnamefont {J.}~\bibnamefont
  {Yu}}, \bibinfo {author} {\bibfnamefont {T.~W.}\ \bibnamefont {Noh}},
  \bibinfo {author} {\bibfnamefont {C.}~\bibnamefont {Kim}}, \bibinfo {author}
  {\bibfnamefont {S.~J.}\ \bibnamefont {Oh}}, \bibinfo {author} {\bibfnamefont
  {J.~H.}\ \bibnamefont {Park}}, \bibinfo {author} {\bibfnamefont
  {V.}~\bibnamefont {Durairaj}}, \bibinfo {author} {\bibfnamefont
  {G.}~\bibnamefont {Cao}}, \ and\ \bibinfo {author} {\bibfnamefont
  {E.}~\bibnamefont {Rotenberg}},\ }\href {\doibase
  10.1103/PhysRevLett.101.076402} {\bibfield  {journal} {\bibinfo  {journal}
  {Physical Review Letters}\ }\textbf {\bibinfo {volume} {101}} (\bibinfo
  {year} {2008}),\ 10.1103/PhysRevLett.101.076402}\BibitemShut {NoStop}%
\bibitem [{\citenamefont {Kim}\ \emph {et~al.}(2009)\citenamefont {Kim},
  \citenamefont {Ohsumi}, \citenamefont {Komesu}, \citenamefont {Sakai},
  \citenamefont {Morita}, \citenamefont {Takagi},\ and\ \citenamefont
  {Arima}}]{kim:09}%
  \BibitemOpen
  \bibfield  {author} {\bibinfo {author} {\bibfnamefont {B.~J.}\ \bibnamefont
  {Kim}}, \bibinfo {author} {\bibfnamefont {H.}~\bibnamefont {Ohsumi}},
  \bibinfo {author} {\bibfnamefont {T.}~\bibnamefont {Komesu}}, \bibinfo
  {author} {\bibfnamefont {S.}~\bibnamefont {Sakai}}, \bibinfo {author}
  {\bibfnamefont {T.}~\bibnamefont {Morita}}, \bibinfo {author} {\bibfnamefont
  {H.}~\bibnamefont {Takagi}}, \ and\ \bibinfo {author} {\bibfnamefont
  {T.}~\bibnamefont {Arima}},\ }\href {\doibase 10.1126/science.1167106}
  {\bibfield  {journal} {\bibinfo  {journal} {Science}\ }\textbf {\bibinfo
  {volume} {323}} (\bibinfo {year} {2009}),\
  10.1126/science.1167106}\BibitemShut {NoStop}%
\bibitem [{\citenamefont {Lado}\ and\ \citenamefont {Pardo}(2015)}]{PardoPRB}%
  \BibitemOpen
  \bibfield  {author} {\bibinfo {author} {\bibfnamefont {J.~L.}\ \bibnamefont
  {Lado}}\ and\ \bibinfo {author} {\bibfnamefont {V.}~\bibnamefont {Pardo}},\
  }\href {\doibase 10.1103/PhysRevB.92.155151} {\bibfield  {journal} {\bibinfo
  {journal} {Physical Review B}\ }\textbf {\bibinfo {volume} {92}} (\bibinfo
  {year} {2015}),\ 10.1103/PhysRevB.92.155151}\BibitemShut {NoStop}%
\bibitem [{\citenamefont {Liu}\ \emph {et~al.}(2016)\citenamefont {Liu},
  \citenamefont {Li}, \citenamefont {Li}, \citenamefont {Liu}, \citenamefont
  {Li}, \citenamefont {Yang}, \citenamefont {Yao}, \citenamefont {Fan},
  \citenamefont {Wan}, \citenamefont {Wang},\ and\ \citenamefont
  {Shen}}]{liu:16}%
  \BibitemOpen
  \bibfield  {author} {\bibinfo {author} {\bibfnamefont {Z.~T.}\ \bibnamefont
  {Liu}}, \bibinfo {author} {\bibfnamefont {M.~Y.}\ \bibnamefont {Li}},
  \bibinfo {author} {\bibfnamefont {Q.~F.}\ \bibnamefont {Li}}, \bibinfo
  {author} {\bibfnamefont {J.~S.}\ \bibnamefont {Liu}}, \bibinfo {author}
  {\bibfnamefont {W.}~\bibnamefont {Li}}, \bibinfo {author} {\bibfnamefont
  {H.~F.}\ \bibnamefont {Yang}}, \bibinfo {author} {\bibfnamefont
  {Q.}~\bibnamefont {Yao}}, \bibinfo {author} {\bibfnamefont {C.~C.}\
  \bibnamefont {Fan}}, \bibinfo {author} {\bibfnamefont {X.~G.}\ \bibnamefont
  {Wan}}, \bibinfo {author} {\bibfnamefont {Z.}~\bibnamefont {Wang}}, \ and\
  \bibinfo {author} {\bibfnamefont {D.~W.}\ \bibnamefont {Shen}},\ }\href
  {\doibase 10.1038/srep30309} {\bibfield  {journal} {\bibinfo  {journal}
  {Scientific Reports}\ } (\bibinfo {year} {2016}),\
  10.1038/srep30309}\BibitemShut {NoStop}%
\bibitem [{\citenamefont {Okada}\ \emph {et~al.}(2013)\citenamefont {Okada},
  \citenamefont {Walkup}, \citenamefont {Lin}, \citenamefont {Dhital},
  \citenamefont {Chang}, \citenamefont {Khadka}, \citenamefont {Zhou},
  \citenamefont {Jeng}, \citenamefont {Paranjape}, \citenamefont {Bansil},
  \citenamefont {Wang}, \citenamefont {Wilson},\ and\ \citenamefont
  {Madhavan}}]{okada:13}%
  \BibitemOpen
  \bibfield  {author} {\bibinfo {author} {\bibfnamefont {Y.}~\bibnamefont
  {Okada}}, \bibinfo {author} {\bibfnamefont {D.}~\bibnamefont {Walkup}},
  \bibinfo {author} {\bibfnamefont {H.}~\bibnamefont {Lin}}, \bibinfo {author}
  {\bibfnamefont {C.}~\bibnamefont {Dhital}}, \bibinfo {author} {\bibfnamefont
  {T.~R.}\ \bibnamefont {Chang}}, \bibinfo {author} {\bibfnamefont
  {S.}~\bibnamefont {Khadka}}, \bibinfo {author} {\bibfnamefont {W.~W.}\
  \bibnamefont {Zhou}}, \bibinfo {author} {\bibfnamefont {H.~T.}\ \bibnamefont
  {Jeng}}, \bibinfo {author} {\bibfnamefont {M.}~\bibnamefont {Paranjape}},
  \bibinfo {author} {\bibfnamefont {A.}~\bibnamefont {Bansil}}, \bibinfo
  {author} {\bibfnamefont {Z.~Q.}\ \bibnamefont {Wang}}, \bibinfo {author}
  {\bibfnamefont {S.~D.}\ \bibnamefont {Wilson}}, \ and\ \bibinfo {author}
  {\bibfnamefont {V.}~\bibnamefont {Madhavan}},\ }\href {\doibase
  10.1038/nmat3653} {\bibfield  {journal} {\bibinfo  {journal} {Nature
  Materials}\ }\textbf {\bibinfo {volume} {12}} (\bibinfo {year} {2013}),\
  10.1038/nmat3653}\BibitemShut {NoStop}%
\bibitem [{\citenamefont {Wang}\ \emph {et~al.}(2013)\citenamefont {Wang},
  \citenamefont {Cao}, \citenamefont {Waugh}, \citenamefont {Park},
  \citenamefont {Qi}, \citenamefont {O.~B.~Korneta},\ and\ \citenamefont
  {Dessau}}]{wang:13}%
  \BibitemOpen
  \bibfield  {author} {\bibinfo {author} {\bibfnamefont {Q.}~\bibnamefont
  {Wang}}, \bibinfo {author} {\bibfnamefont {Y.}~\bibnamefont {Cao}}, \bibinfo
  {author} {\bibfnamefont {J.~A.}\ \bibnamefont {Waugh}}, \bibinfo {author}
  {\bibfnamefont {S.~R.}\ \bibnamefont {Park}}, \bibinfo {author}
  {\bibfnamefont {T.~F.}\ \bibnamefont {Qi}}, \bibinfo {author} {\bibfnamefont
  {G.~C.}\ \bibnamefont {O.~B.~Korneta}}, \ and\ \bibinfo {author}
  {\bibfnamefont {D.~S.}\ \bibnamefont {Dessau}},\ }\href {\doibase
  10.1103/PhysRevB.87.245109} {\bibfield  {journal} {\bibinfo  {journal}
  {Physical Review B}\ }\textbf {\bibinfo {volume} {87}} (\bibinfo {year}
  {2013}),\ 10.1103/PhysRevB.87.245109}\BibitemShut {NoStop}%
\bibitem [{\citenamefont {Wojek}\ \emph {et~al.}(2012)\citenamefont {Wojek},
  \citenamefont {Berntsen}, \citenamefont {Boseggia}, \citenamefont
  {Boothroyd}, \citenamefont {Prabhakaran}, \citenamefont {McMorrow},
  \citenamefont {Rønnow}, \citenamefont {Chang}, ,\ and\ \citenamefont
  {Tjernberg}}]{wojek:12}%
  \BibitemOpen
  \bibfield  {author} {\bibinfo {author} {\bibfnamefont {B.~M.}\ \bibnamefont
  {Wojek}}, \bibinfo {author} {\bibfnamefont {M.~H.}\ \bibnamefont {Berntsen}},
  \bibinfo {author} {\bibfnamefont {S.}~\bibnamefont {Boseggia}}, \bibinfo
  {author} {\bibfnamefont {A.~T.}\ \bibnamefont {Boothroyd}}, \bibinfo {author}
  {\bibfnamefont {D.}~\bibnamefont {Prabhakaran}}, \bibinfo {author}
  {\bibfnamefont {D.~F.}\ \bibnamefont {McMorrow}}, \bibinfo {author}
  {\bibfnamefont {H.~M.}\ \bibnamefont {Rønnow}}, \bibinfo {author}
  {\bibfnamefont {J.}~\bibnamefont {Chang}}, , \ and\ \bibinfo {author}
  {\bibfnamefont {O.}~\bibnamefont {Tjernberg}},\ }\href {\doibase
  10.1088/0953-8984/24/41/415602} {\bibfield  {journal} {\bibinfo  {journal}
  {Journal of Physics: Condensed Matter}\ }\textbf {\bibinfo {volume} {24}}
  (\bibinfo {year} {2012}),\ 10.1088/0953-8984/24/41/415602}\BibitemShut
  {NoStop}%
\bibitem [{\citenamefont {Nichols}\ \emph {et~al.}(2013)\citenamefont
  {Nichols}, \citenamefont {Terzic}, \citenamefont {Bittle}, \citenamefont
  {Korneta}, \citenamefont {Long}, \citenamefont {Brill}, \citenamefont {Cao},\
  and\ \citenamefont {Seo}}]{nichols:13}%
  \BibitemOpen
  \bibfield  {author} {\bibinfo {author} {\bibfnamefont {J.}~\bibnamefont
  {Nichols}}, \bibinfo {author} {\bibfnamefont {J.}~\bibnamefont {Terzic}},
  \bibinfo {author} {\bibfnamefont {E.~G.}\ \bibnamefont {Bittle}}, \bibinfo
  {author} {\bibfnamefont {O.~B.}\ \bibnamefont {Korneta}}, \bibinfo {author}
  {\bibfnamefont {L.~E.~D.}\ \bibnamefont {Long}}, \bibinfo {author}
  {\bibfnamefont {J.~W.}\ \bibnamefont {Brill}}, \bibinfo {author}
  {\bibfnamefont {G.}~\bibnamefont {Cao}}, \ and\ \bibinfo {author}
  {\bibfnamefont {S.~S.~A.}\ \bibnamefont {Seo}},\ }\href {\doibase
  10.1063/1.4801877} {\bibfield  {journal} {\bibinfo  {journal} {Applied
  Physics Letters}\ }\textbf {\bibinfo {volume} {102}} (\bibinfo {year}
  {2013}),\ 10.1063/1.4801877}\BibitemShut {NoStop}%
\bibitem [{\citenamefont {Rayan~Serrao}\ \emph {et~al.}(2013)\citenamefont
  {Rayan~Serrao}, \citenamefont {Liu}, \citenamefont {Heron}, \citenamefont
  {Singh-Bhalla}, \citenamefont {Yadav}, \citenamefont {Suresha}, \citenamefont
  {Paull}, \citenamefont {Yi}, \citenamefont {Chu}, \citenamefont {Trassin},
  \citenamefont {Vishwanath}, \citenamefont {Arenholz}, \citenamefont
  {Frontera}, \citenamefont {\ifmmode~\check{Z}\else \v{Z}\fi{}elezn\'y},
  \citenamefont {Jungwirth}, \citenamefont {Marti},\ and\ \citenamefont
  {Ramesh}}]{serrao:13}%
  \BibitemOpen
  \bibfield  {author} {\bibinfo {author} {\bibfnamefont {C.}~\bibnamefont
  {Rayan~Serrao}}, \bibinfo {author} {\bibfnamefont {J.}~\bibnamefont {Liu}},
  \bibinfo {author} {\bibfnamefont {J.~T.}\ \bibnamefont {Heron}}, \bibinfo
  {author} {\bibfnamefont {G.}~\bibnamefont {Singh-Bhalla}}, \bibinfo {author}
  {\bibfnamefont {A.}~\bibnamefont {Yadav}}, \bibinfo {author} {\bibfnamefont
  {S.~J.}\ \bibnamefont {Suresha}}, \bibinfo {author} {\bibfnamefont {R.~J.}\
  \bibnamefont {Paull}}, \bibinfo {author} {\bibfnamefont {D.}~\bibnamefont
  {Yi}}, \bibinfo {author} {\bibfnamefont {J.-H.}\ \bibnamefont {Chu}},
  \bibinfo {author} {\bibfnamefont {M.}~\bibnamefont {Trassin}}, \bibinfo
  {author} {\bibfnamefont {A.}~\bibnamefont {Vishwanath}}, \bibinfo {author}
  {\bibfnamefont {E.}~\bibnamefont {Arenholz}}, \bibinfo {author}
  {\bibfnamefont {C.}~\bibnamefont {Frontera}}, \bibinfo {author}
  {\bibfnamefont {J.}~\bibnamefont {\ifmmode~\check{Z}\else
  \v{Z}\fi{}elezn\'y}}, \bibinfo {author} {\bibfnamefont {T.}~\bibnamefont
  {Jungwirth}}, \bibinfo {author} {\bibfnamefont {X.}~\bibnamefont {Marti}}, \
  and\ \bibinfo {author} {\bibfnamefont {R.}~\bibnamefont {Ramesh}},\ }\href
  {\doibase 10.1103/PhysRevB.87.085121} {\bibfield  {journal} {\bibinfo
  {journal} {Physical Review B}\ }\textbf {\bibinfo {volume} {87}} (\bibinfo
  {year} {2013}),\ 10.1103/PhysRevB.87.085121}\BibitemShut {NoStop}%
\bibitem [{\citenamefont {Gruenewald}\ \emph {et~al.}(2014)\citenamefont
  {Gruenewald}, \citenamefont {Nichols}, \citenamefont {Terzic}, \citenamefont
  {Cao}, \citenamefont {Brill},\ and\ \citenamefont {Seo}}]{gruenewald:14}%
  \BibitemOpen
  \bibfield  {author} {\bibinfo {author} {\bibfnamefont {J.~H.}\ \bibnamefont
  {Gruenewald}}, \bibinfo {author} {\bibfnamefont {J.}~\bibnamefont {Nichols}},
  \bibinfo {author} {\bibfnamefont {J.}~\bibnamefont {Terzic}}, \bibinfo
  {author} {\bibfnamefont {G.}~\bibnamefont {Cao}}, \bibinfo {author}
  {\bibfnamefont {J.~W.}\ \bibnamefont {Brill}}, \ and\ \bibinfo {author}
  {\bibfnamefont {S.~S.~A.}\ \bibnamefont {Seo}},\ }\href {\doibase
  10.1557/jmr.2014.288} {\bibfield  {journal} {\bibinfo  {journal} {Journal of
  Materials Research}\ }\textbf {\bibinfo {volume} {29}} (\bibinfo {year}
  {2014}),\ 10.1557/jmr.2014.288}\BibitemShut {NoStop}%
\bibitem [{\citenamefont {Moon}\ \emph {et~al.}(2008)\citenamefont {Moon},
  \citenamefont {Jin}, \citenamefont {Kim}, \citenamefont {Choi}, \citenamefont
  {Lee}, \citenamefont {Yu}, \citenamefont {Cao}, \citenamefont {Sumi},
  \citenamefont {Funakubo}, \citenamefont {Bernhard},\ and\ \citenamefont
  {Noh}}]{moon:08}%
  \BibitemOpen
  \bibfield  {author} {\bibinfo {author} {\bibfnamefont {S.~J.}\ \bibnamefont
  {Moon}}, \bibinfo {author} {\bibfnamefont {H.}~\bibnamefont {Jin}}, \bibinfo
  {author} {\bibfnamefont {K.~W.}\ \bibnamefont {Kim}}, \bibinfo {author}
  {\bibfnamefont {W.~S.}\ \bibnamefont {Choi}}, \bibinfo {author}
  {\bibfnamefont {Y.~S.}\ \bibnamefont {Lee}}, \bibinfo {author} {\bibfnamefont
  {J.}~\bibnamefont {Yu}}, \bibinfo {author} {\bibfnamefont {G.}~\bibnamefont
  {Cao}}, \bibinfo {author} {\bibfnamefont {A.}~\bibnamefont {Sumi}}, \bibinfo
  {author} {\bibfnamefont {H.}~\bibnamefont {Funakubo}}, \bibinfo {author}
  {\bibfnamefont {C.}~\bibnamefont {Bernhard}}, \ and\ \bibinfo {author}
  {\bibfnamefont {T.~W.}\ \bibnamefont {Noh}},\ }\href {\doibase
  10.1103/PhysRevLett.101.226402} {\bibfield  {journal} {\bibinfo  {journal}
  {Physical Review Letters}\ }\textbf {\bibinfo {volume} {101}} (\bibinfo
  {year} {2008}),\ 10.1103/PhysRevLett.101.226402}\BibitemShut {NoStop}%
\bibitem [{\citenamefont {Pallecchi}\ \emph {et~al.}(2016)\citenamefont
  {Pallecchi}, \citenamefont {Buscaglia}, \citenamefont {Buscaglia},
  \citenamefont {Gilioli}, \citenamefont {Lamura}, \citenamefont {Telesio},
  \citenamefont {Cimberle},\ and\ \citenamefont {Marre}}]{pallecchi:16}%
  \BibitemOpen
  \bibfield  {author} {\bibinfo {author} {\bibfnamefont {I.}~\bibnamefont
  {Pallecchi}}, \bibinfo {author} {\bibfnamefont {M.~T.}\ \bibnamefont
  {Buscaglia}}, \bibinfo {author} {\bibfnamefont {V.}~\bibnamefont
  {Buscaglia}}, \bibinfo {author} {\bibfnamefont {E.}~\bibnamefont {Gilioli}},
  \bibinfo {author} {\bibfnamefont {G.}~\bibnamefont {Lamura}}, \bibinfo
  {author} {\bibfnamefont {F.}~\bibnamefont {Telesio}}, \bibinfo {author}
  {\bibfnamefont {M.~R.}\ \bibnamefont {Cimberle}}, \ and\ \bibinfo {author}
  {\bibfnamefont {D.}~\bibnamefont {Marre}},\ }\href {\doibase
  10.1088/0953-8984/28/6/065601} {\bibfield  {journal} {\bibinfo  {journal}
  {Journal of Physics--Condensed Matter}\ }\textbf {\bibinfo {volume} {28}}
  (\bibinfo {year} {2016}),\ 10.1088/0953-8984/28/6/065601}\BibitemShut
  {NoStop}%
\bibitem [{\citenamefont {Seo}\ \emph {et~al.}(2016)\citenamefont {Seo},
  \citenamefont {Nichols}, \citenamefont {Hwang}, \citenamefont {Terzic},
  \citenamefont {Gruenewald}, \citenamefont {Souri}, \citenamefont {Thompson},
  \citenamefont {Connell},\ and\ \citenamefont {Cao}}]{seo:16}%
  \BibitemOpen
  \bibfield  {author} {\bibinfo {author} {\bibfnamefont {S.~S.~A.}\
  \bibnamefont {Seo}}, \bibinfo {author} {\bibfnamefont {J.}~\bibnamefont
  {Nichols}}, \bibinfo {author} {\bibfnamefont {J.}~\bibnamefont {Hwang}},
  \bibinfo {author} {\bibfnamefont {J.}~\bibnamefont {Terzic}}, \bibinfo
  {author} {\bibfnamefont {J.~H.}\ \bibnamefont {Gruenewald}}, \bibinfo
  {author} {\bibfnamefont {M.}~\bibnamefont {Souri}}, \bibinfo {author}
  {\bibfnamefont {J.}~\bibnamefont {Thompson}}, \bibinfo {author}
  {\bibfnamefont {J.~G.}\ \bibnamefont {Connell}}, \ and\ \bibinfo {author}
  {\bibfnamefont {G.}~\bibnamefont {Cao}},\ }\href {\doibase 10.1063/1.4967450}
  {\bibfield  {journal} {\bibinfo  {journal} {Applied Physics Letters}\
  }\textbf {\bibinfo {volume} {109}} (\bibinfo {year} {2016}),\
  10.1063/1.4967450}\BibitemShut {NoStop}%
\bibitem [{\citenamefont {Lee}\ \emph {et~al.}(2014)\citenamefont {Lee},
  \citenamefont {Luo}, \citenamefont {Tung}, \citenamefont {Chang},
  \citenamefont {Luo}, \citenamefont {Malshe}, \citenamefont {Gadre},
  \citenamefont {Bhattacharya}, \citenamefont {Nakhmanson}, \citenamefont
  {Eastman}, \citenamefont {Hong}, \citenamefont {Jellinek}, \citenamefont
  {Morgan}, \citenamefont {Fong},\ and\ \citenamefont {Freeland}}]{lee:14}%
  \BibitemOpen
  \bibfield  {author} {\bibinfo {author} {\bibfnamefont {J.~H.}\ \bibnamefont
  {Lee}}, \bibinfo {author} {\bibfnamefont {G.}~\bibnamefont {Luo}}, \bibinfo
  {author} {\bibfnamefont {I.~C.}\ \bibnamefont {Tung}}, \bibinfo {author}
  {\bibfnamefont {S.~H.}\ \bibnamefont {Chang}}, \bibinfo {author}
  {\bibfnamefont {Z.}~\bibnamefont {Luo}}, \bibinfo {author} {\bibfnamefont
  {M.}~\bibnamefont {Malshe}}, \bibinfo {author} {\bibfnamefont
  {M.}~\bibnamefont {Gadre}}, \bibinfo {author} {\bibfnamefont
  {A.}~\bibnamefont {Bhattacharya}}, \bibinfo {author} {\bibfnamefont {S.~M.}\
  \bibnamefont {Nakhmanson}}, \bibinfo {author} {\bibfnamefont {J.~A.}\
  \bibnamefont {Eastman}}, \bibinfo {author} {\bibfnamefont {H.}~\bibnamefont
  {Hong}}, \bibinfo {author} {\bibfnamefont {J.}~\bibnamefont {Jellinek}},
  \bibinfo {author} {\bibfnamefont {D.}~\bibnamefont {Morgan}}, \bibinfo
  {author} {\bibfnamefont {D.~D.}\ \bibnamefont {Fong}}, \ and\ \bibinfo
  {author} {\bibfnamefont {J.}~\bibnamefont {Freeland}},\ }\href {\doibase
  10.1038/nmat4039} {\bibfield  {journal} {\bibinfo  {journal} {Nature
  Materials}\ }\textbf {\bibinfo {volume} {13}} (\bibinfo {year} {2014}),\
  10.1038/nmat4039}\BibitemShut {NoStop}%
\bibitem [{\citenamefont {Nie}\ \emph {et~al.}(2014)\citenamefont {Nie},
  \citenamefont {Zhu}, \citenamefont {Lee}, \citenamefont {Kourkoutis},
  \citenamefont {Mundy}, \citenamefont {Junquera}, \citenamefont {Ghosez},
  \citenamefont {Baek}, \citenamefont {Sung}, \citenamefont {Xi}, \citenamefont
  {Shen}, \citenamefont {Muller},\ and\ \citenamefont {Schlom}}]{nie:14}%
  \BibitemOpen
  \bibfield  {author} {\bibinfo {author} {\bibfnamefont {Y.~F.}\ \bibnamefont
  {Nie}}, \bibinfo {author} {\bibfnamefont {Y.}~\bibnamefont {Zhu}}, \bibinfo
  {author} {\bibfnamefont {C.~H.}\ \bibnamefont {Lee}}, \bibinfo {author}
  {\bibfnamefont {L.~F.}\ \bibnamefont {Kourkoutis}}, \bibinfo {author}
  {\bibfnamefont {J.~A.}\ \bibnamefont {Mundy}}, \bibinfo {author}
  {\bibfnamefont {J.}~\bibnamefont {Junquera}}, \bibinfo {author}
  {\bibfnamefont {P.}~\bibnamefont {Ghosez}}, \bibinfo {author} {\bibfnamefont
  {D.~J.}\ \bibnamefont {Baek}}, \bibinfo {author} {\bibfnamefont
  {S.}~\bibnamefont {Sung}}, \bibinfo {author} {\bibfnamefont {X.~X.}\
  \bibnamefont {Xi}}, \bibinfo {author} {\bibfnamefont {K.~M.}\ \bibnamefont
  {Shen}}, \bibinfo {author} {\bibfnamefont {D.~A.}\ \bibnamefont {Muller}}, \
  and\ \bibinfo {author} {\bibfnamefont {D.~G.}\ \bibnamefont {Schlom}},\
  }\href {\doibase 10.1038/ncomms5530} {\bibfield  {journal} {\bibinfo
  {journal} {Nature Communications}\ }\textbf {\bibinfo {volume} {5}} (\bibinfo
  {year} {2014}),\ 10.1038/ncomms5530}\BibitemShut {NoStop}%
\bibitem [{sup()}]{supplementary}%
  \BibitemOpen
  \href@noop {} {}\bibinfo {note} {See supplementary material for a detailed
  experimental description, additional figures, and a comment on the impact of
  volatile IrO$_2$}\BibitemShut {NoStop}%
\bibitem [{\citenamefont {Nishio}\ \emph {et~al.}(2016)\citenamefont {Nishio},
  \citenamefont {Hwang},\ and\ \citenamefont {Hikita}}]{nishio:16}%
  \BibitemOpen
  \bibfield  {author} {\bibinfo {author} {\bibfnamefont {K.}~\bibnamefont
  {Nishio}}, \bibinfo {author} {\bibfnamefont {H.~Y.}\ \bibnamefont {Hwang}}, \
  and\ \bibinfo {author} {\bibfnamefont {Y.}~\bibnamefont {Hikita}},\ }\href
  {\doibase 10.1063/1.4943519} {\bibfield  {journal} {\bibinfo  {journal} {APL
  Materials}\ }\textbf {\bibinfo {volume} {4}} (\bibinfo {year} {2016}),\
  10.1063/1.4943519}\BibitemShut {NoStop}%
\bibitem [{\citenamefont {Liu}\ \emph {et~al.}(2017)\citenamefont {Liu},
  \citenamefont {Cao}, \citenamefont {Pal}, \citenamefont {Middey},
  \citenamefont {Kareev}, \citenamefont {Choi}, \citenamefont {Shafer},
  \citenamefont {Haskel}, \citenamefont {Arenholz},\ and\ \citenamefont
  {Chakhalian}}]{liu:17}%
  \BibitemOpen
  \bibfield  {author} {\bibinfo {author} {\bibfnamefont {X.}~\bibnamefont
  {Liu}}, \bibinfo {author} {\bibfnamefont {Y.}~\bibnamefont {Cao}}, \bibinfo
  {author} {\bibfnamefont {B.}~\bibnamefont {Pal}}, \bibinfo {author}
  {\bibfnamefont {S.}~\bibnamefont {Middey}}, \bibinfo {author} {\bibfnamefont
  {M.}~\bibnamefont {Kareev}}, \bibinfo {author} {\bibfnamefont
  {Y.}~\bibnamefont {Choi}}, \bibinfo {author} {\bibfnamefont {P.}~\bibnamefont
  {Shafer}}, \bibinfo {author} {\bibfnamefont {D.}~\bibnamefont {Haskel}},
  \bibinfo {author} {\bibfnamefont {E.}~\bibnamefont {Arenholz}}, \ and\
  \bibinfo {author} {\bibfnamefont {J.}~\bibnamefont {Chakhalian}},\ }\href
  {\doibase 10.1103/PhysRevMaterials.1.075004} {\bibfield  {journal} {\bibinfo
  {journal} {Physical Review Materials}\ }\textbf {\bibinfo {volume} {1}}
  (\bibinfo {year} {2017}),\ 10.1103/PhysRevMaterials.1.075004}\BibitemShut
  {NoStop}%
\bibitem [{\citenamefont {Huang}\ \emph {et~al.}(1994)\citenamefont {Huang},
  \citenamefont {Soubeyroux}, \citenamefont {Chmaissem}, \citenamefont {Sora},
  \citenamefont {Santoro}, \citenamefont {Cava}, \citenamefont {Krajewski},\
  and\ \citenamefont {Jr}}]{huang:94}%
  \BibitemOpen
  \bibfield  {author} {\bibinfo {author} {\bibfnamefont {Q.}~\bibnamefont
  {Huang}}, \bibinfo {author} {\bibfnamefont {J.~L.}\ \bibnamefont
  {Soubeyroux}}, \bibinfo {author} {\bibfnamefont {O.}~\bibnamefont
  {Chmaissem}}, \bibinfo {author} {\bibfnamefont {I.~N.}\ \bibnamefont {Sora}},
  \bibinfo {author} {\bibfnamefont {A.}~\bibnamefont {Santoro}}, \bibinfo
  {author} {\bibfnamefont {R.~J.}\ \bibnamefont {Cava}}, \bibinfo {author}
  {\bibfnamefont {J.~J.}\ \bibnamefont {Krajewski}}, \ and\ \bibinfo {author}
  {\bibfnamefont {W.~F.~P.}\ \bibnamefont {Jr}},\ }\href {\doibase
  10.1006/jssc.1994.1316} {\bibfield  {journal} {\bibinfo  {journal} {Journal
  of Solid State Chemistry}\ }\textbf {\bibinfo {volume} {112}} (\bibinfo
  {year} {1994}),\ 10.1006/jssc.1994.1316}\BibitemShut {NoStop}%
\bibitem [{\citenamefont {Hogan}\ \emph {et~al.}(2016)\citenamefont {Hogan},
  \citenamefont {Bjaalie}, \citenamefont {Zhao}, \citenamefont {Belvin},
  \citenamefont {Wang}, \citenamefont {de~Walle}, \citenamefont {Hsieh},\ and\
  \citenamefont {Wilson}}]{hogan:16}%
  \BibitemOpen
  \bibfield  {author} {\bibinfo {author} {\bibfnamefont {T.}~\bibnamefont
  {Hogan}}, \bibinfo {author} {\bibfnamefont {L.}~\bibnamefont {Bjaalie}},
  \bibinfo {author} {\bibfnamefont {L.}~\bibnamefont {Zhao}}, \bibinfo {author}
  {\bibfnamefont {C.}~\bibnamefont {Belvin}}, \bibinfo {author} {\bibfnamefont
  {X.}~\bibnamefont {Wang}}, \bibinfo {author} {\bibfnamefont {C.~G.~V.}\
  \bibnamefont {de~Walle}}, \bibinfo {author} {\bibfnamefont {D.}~\bibnamefont
  {Hsieh}}, \ and\ \bibinfo {author} {\bibfnamefont {S.~D.}\ \bibnamefont
  {Wilson}},\ }\href {\doibase 10.1103/PhysRevB.93.134110} {\bibfield
  {journal} {\bibinfo  {journal} {Physical Review B}\ }\textbf {\bibinfo
  {volume} {93}} (\bibinfo {year} {2016}),\
  10.1103/PhysRevB.93.134110}\BibitemShut {NoStop}%
\bibitem [{\citenamefont {Zhao}\ \emph {et~al.}(2008)\citenamefont {Zhao},
  \citenamefont {Yang}, \citenamefont {Yu}, \citenamefont {Li}, \citenamefont
  {Yu}, \citenamefont {Fang}, \citenamefont {Chen},\ and\ \citenamefont
  {Jin}}]{zhao:08}%
  \BibitemOpen
  \bibfield  {author} {\bibinfo {author} {\bibfnamefont {J.~G.}\ \bibnamefont
  {Zhao}}, \bibinfo {author} {\bibfnamefont {L.~X.}\ \bibnamefont {Yang}},
  \bibinfo {author} {\bibfnamefont {Y.}~\bibnamefont {Yu}}, \bibinfo {author}
  {\bibfnamefont {F.~Y.}\ \bibnamefont {Li}}, \bibinfo {author} {\bibfnamefont
  {R.~C.}\ \bibnamefont {Yu}}, \bibinfo {author} {\bibfnamefont
  {Z.}~\bibnamefont {Fang}}, \bibinfo {author} {\bibfnamefont {L.~C.}\
  \bibnamefont {Chen}}, \ and\ \bibinfo {author} {\bibfnamefont {C.~Q.}\
  \bibnamefont {Jin}},\ }\href {\doibase 10.1063/1.2908879} {\bibfield
  {journal} {\bibinfo  {journal} {Journal of Applied Physics}\ }\textbf
  {\bibinfo {volume} {103}} (\bibinfo {year} {2008}),\
  10.1063/1.2908879}\BibitemShut {NoStop}%
\bibitem [{\citenamefont {Puggioni}\ and\ \citenamefont
  {Rondinelli}(2016)}]{puggioni:16}%
  \BibitemOpen
  \bibfield  {author} {\bibinfo {author} {\bibfnamefont {D.}~\bibnamefont
  {Puggioni}}\ and\ \bibinfo {author} {\bibfnamefont {J.~M.}\ \bibnamefont
  {Rondinelli}},\ }\href {\doibase 10.1063/1.4942651} {\bibfield  {journal}
  {\bibinfo  {journal} {Journal of Applied Physics}\ }\textbf {\bibinfo
  {volume} {119}} (\bibinfo {year} {2016}),\ 10.1063/1.4942651}\BibitemShut
  {NoStop}%
\bibitem [{\citenamefont {Wicklein}\ \emph {et~al.}(2012)\citenamefont
  {Wicklein}, \citenamefont {Sambri}, \citenamefont {Amoruso}, \citenamefont
  {Wang}, \citenamefont {Bruzzese}, \citenamefont {Koehl},\ and\ \citenamefont
  {Dittmann}}]{wicklein:12}%
  \BibitemOpen
  \bibfield  {author} {\bibinfo {author} {\bibfnamefont {S.}~\bibnamefont
  {Wicklein}}, \bibinfo {author} {\bibfnamefont {A.}~\bibnamefont {Sambri}},
  \bibinfo {author} {\bibfnamefont {S.}~\bibnamefont {Amoruso}}, \bibinfo
  {author} {\bibfnamefont {X.}~\bibnamefont {Wang}}, \bibinfo {author}
  {\bibfnamefont {R.}~\bibnamefont {Bruzzese}}, \bibinfo {author}
  {\bibfnamefont {A.}~\bibnamefont {Koehl}}, \ and\ \bibinfo {author}
  {\bibfnamefont {R.}~\bibnamefont {Dittmann}},\ }\href {\doibase
  10.1063/1.4754112} {\bibfield  {journal} {\bibinfo  {journal} {Applied
  Physics Letters}\ }\textbf {\bibinfo {volume} {101}} (\bibinfo {year}
  {2012}),\ 10.1063/1.4754112}\BibitemShut {NoStop}%
\bibitem [{\citenamefont {Schraknepper}\ \emph {et~al.}(2016)\citenamefont
  {Schraknepper}, \citenamefont {Baumer}, \citenamefont {Gunkel}, \citenamefont
  {Dittmann},\ and\ \citenamefont {Souza}}]{schraknepper:16}%
  \BibitemOpen
  \bibfield  {author} {\bibinfo {author} {\bibfnamefont {H.}~\bibnamefont
  {Schraknepper}}, \bibinfo {author} {\bibfnamefont {C.}~\bibnamefont
  {Baumer}}, \bibinfo {author} {\bibfnamefont {F.}~\bibnamefont {Gunkel}},
  \bibinfo {author} {\bibfnamefont {R.}~\bibnamefont {Dittmann}}, \ and\
  \bibinfo {author} {\bibfnamefont {R.~A.~D.}\ \bibnamefont {Souza}},\ }\href
  {\doibase 10.1063/1.4972996} {\bibfield  {journal} {\bibinfo  {journal} {APL
  Materials}\ }\textbf {\bibinfo {volume} {4}} (\bibinfo {year} {2016}),\
  10.1063/1.4972996}\BibitemShut {NoStop}%
\bibitem [{\citenamefont {Packwood}\ \emph {et~al.}(2013)\citenamefont
  {Packwood}, \citenamefont {Shiraki},\ and\ \citenamefont
  {Hitosugi}}]{packwood:13}%
  \BibitemOpen
  \bibfield  {author} {\bibinfo {author} {\bibfnamefont {D.~M.}\ \bibnamefont
  {Packwood}}, \bibinfo {author} {\bibfnamefont {S.}~\bibnamefont {Shiraki}}, \
  and\ \bibinfo {author} {\bibfnamefont {T.}~\bibnamefont {Hitosugi}},\ }\href
  {\doibase 10.1103/PhysRevLett.111.036101} {\bibfield  {journal} {\bibinfo
  {journal} {Physical Review Letters}\ }\textbf {\bibinfo {volume} {111}}
  (\bibinfo {year} {2013}),\ 10.1103/PhysRevLett.111.036101}\BibitemShut
  {NoStop}%
\bibitem [{\citenamefont {Sambri}\ \emph {et~al.}(2016)\citenamefont {Sambri},
  \citenamefont {Aruta}, \citenamefont {Gennaro}, \citenamefont {Wang},
  \citenamefont {di~Uccio}, \citenamefont {Granozio},\ and\ \citenamefont
  {Amoruso}}]{sambri:16}%
  \BibitemOpen
  \bibfield  {author} {\bibinfo {author} {\bibfnamefont {A.}~\bibnamefont
  {Sambri}}, \bibinfo {author} {\bibfnamefont {C.}~\bibnamefont {Aruta}},
  \bibinfo {author} {\bibfnamefont {E.~D.}\ \bibnamefont {Gennaro}}, \bibinfo
  {author} {\bibfnamefont {X.}~\bibnamefont {Wang}}, \bibinfo {author}
  {\bibfnamefont {U.~S.}\ \bibnamefont {di~Uccio}}, \bibinfo {author}
  {\bibfnamefont {F.~M.}\ \bibnamefont {Granozio}}, \ and\ \bibinfo {author}
  {\bibfnamefont {S.}~\bibnamefont {Amoruso}},\ }\href {\doibase
  10.1063/1.4943589} {\bibfield  {journal} {\bibinfo  {journal} {Journal of
  Applied Physics}\ }\textbf {\bibinfo {volume} {119}} (\bibinfo {year}
  {2016}),\ 10.1063/1.4943589}\BibitemShut {NoStop}%
\bibitem [{\citenamefont {Groenen}\ \emph {et~al.}(2015)\citenamefont
  {Groenen}, \citenamefont {Smit}, \citenamefont {Orsel}, \citenamefont
  {Vailionis}, \citenamefont {Bastiaens}, \citenamefont {Huijben},
  \citenamefont {Boller}, \citenamefont {Rijnders},\ and\ \citenamefont
  {Koster}}]{groenen:15}%
  \BibitemOpen
  \bibfield  {author} {\bibinfo {author} {\bibfnamefont {R.}~\bibnamefont
  {Groenen}}, \bibinfo {author} {\bibfnamefont {J.}~\bibnamefont {Smit}},
  \bibinfo {author} {\bibfnamefont {K.}~\bibnamefont {Orsel}}, \bibinfo
  {author} {\bibfnamefont {A.}~\bibnamefont {Vailionis}}, \bibinfo {author}
  {\bibfnamefont {B.}~\bibnamefont {Bastiaens}}, \bibinfo {author}
  {\bibfnamefont {M.}~\bibnamefont {Huijben}}, \bibinfo {author} {\bibfnamefont
  {K.}~\bibnamefont {Boller}}, \bibinfo {author} {\bibfnamefont
  {G.}~\bibnamefont {Rijnders}}, \ and\ \bibinfo {author} {\bibfnamefont
  {G.}~\bibnamefont {Koster}},\ }\href {\doibase 10.1063/1.4926933} {\bibfield
  {journal} {\bibinfo  {journal} {APL Materials}\ }\textbf {\bibinfo {volume}
  {3}} (\bibinfo {year} {2015}),\ 10.1063/1.4926933}\BibitemShut {NoStop}%
\bibitem [{\citenamefont {Tselev}\ \emph {et~al.}(2016)\citenamefont {Tselev},
  \citenamefont {Vasudevan}, \citenamefont {Gianfrancesco}, \citenamefont
  {Qiao}, \citenamefont {Meyer}, \citenamefont {Lee}, \citenamefont
  {Biegalski}, \citenamefont {Baddorf},\ and\ \citenamefont
  {Kalinin}}]{tselev:16}%
  \BibitemOpen
  \bibfield  {author} {\bibinfo {author} {\bibfnamefont {A.}~\bibnamefont
  {Tselev}}, \bibinfo {author} {\bibfnamefont {R.~K.}\ \bibnamefont
  {Vasudevan}}, \bibinfo {author} {\bibfnamefont {A.~G.}\ \bibnamefont
  {Gianfrancesco}}, \bibinfo {author} {\bibfnamefont {L.}~\bibnamefont {Qiao}},
  \bibinfo {author} {\bibfnamefont {T.~L.}\ \bibnamefont {Meyer}}, \bibinfo
  {author} {\bibfnamefont {H.~N.}\ \bibnamefont {Lee}}, \bibinfo {author}
  {\bibfnamefont {M.~D.}\ \bibnamefont {Biegalski}}, \bibinfo {author}
  {\bibfnamefont {A.~P.}\ \bibnamefont {Baddorf}}, \ and\ \bibinfo {author}
  {\bibfnamefont {S.~V.}\ \bibnamefont {Kalinin}},\ }\href {\doibase
  10.1021/acs.cgd.5b01826} {\bibfield  {journal} {\bibinfo  {journal} {Crystal
  Growth and Design}\ }\textbf {\bibinfo {volume} {16}} (\bibinfo {year}
  {2016}),\ 10.1021/acs.cgd.5b01826}\BibitemShut {NoStop}%
\bibitem [{\citenamefont {Brune}\ \emph {et~al.}(1995)\citenamefont {Brune},
  \citenamefont {Bromann}, \citenamefont {Röder}, \citenamefont {Kern},
  \citenamefont {Jacobsen}, \citenamefont {Stoltze}, \citenamefont {Jacobsen},\
  and\ \citenamefont {Norskov}}]{brune:95}%
  \BibitemOpen
  \bibfield  {author} {\bibinfo {author} {\bibfnamefont {H.}~\bibnamefont
  {Brune}}, \bibinfo {author} {\bibfnamefont {K.}~\bibnamefont {Bromann}},
  \bibinfo {author} {\bibfnamefont {H.}~\bibnamefont {Röder}}, \bibinfo
  {author} {\bibfnamefont {K.}~\bibnamefont {Kern}}, \bibinfo {author}
  {\bibfnamefont {J.}~\bibnamefont {Jacobsen}}, \bibinfo {author}
  {\bibfnamefont {P.}~\bibnamefont {Stoltze}}, \bibinfo {author} {\bibfnamefont
  {K.}~\bibnamefont {Jacobsen}}, \ and\ \bibinfo {author} {\bibfnamefont
  {J.}~\bibnamefont {Norskov}},\ }\href {\doibase 10.1103/PhysRevB.52.R14380}
  {\bibfield  {journal} {\bibinfo  {journal} {Physical Review B}\ }\textbf
  {\bibinfo {volume} {52}} (\bibinfo {year} {1995}),\
  10.1103/PhysRevB.52.R14380}\BibitemShut {NoStop}%
\bibitem [{\citenamefont {Schroeder}\ and\ \citenamefont
  {Wolf}(1997)}]{schroeder:97}%
  \BibitemOpen
  \bibfield  {author} {\bibinfo {author} {\bibfnamefont {M.}~\bibnamefont
  {Schroeder}}\ and\ \bibinfo {author} {\bibfnamefont {D.~E.}\ \bibnamefont
  {Wolf}},\ }\href {\doibase 10.1016/S0039-6028(96)01250-2} {\bibfield
  {journal} {\bibinfo  {journal} {Surface Science}\ }\textbf {\bibinfo {volume}
  {375}} (\bibinfo {year} {1997}),\ 10.1016/S0039-6028(96)01250-2}\BibitemShut
  {NoStop}%
\bibitem [{\citenamefont {Xu}\ \emph {et~al.}(2014)\citenamefont {Xu},
  \citenamefont {Wicklein}, \citenamefont {Sambri}, \citenamefont {Amoruso},
  \citenamefont {Moors},\ and\ \citenamefont {Dittmann}}]{xu:14}%
  \BibitemOpen
  \bibfield  {author} {\bibinfo {author} {\bibfnamefont {C.}~\bibnamefont
  {Xu}}, \bibinfo {author} {\bibfnamefont {S.}~\bibnamefont {Wicklein}},
  \bibinfo {author} {\bibfnamefont {A.}~\bibnamefont {Sambri}}, \bibinfo
  {author} {\bibfnamefont {S.}~\bibnamefont {Amoruso}}, \bibinfo {author}
  {\bibfnamefont {M.}~\bibnamefont {Moors}}, \ and\ \bibinfo {author}
  {\bibfnamefont {R.}~\bibnamefont {Dittmann}},\ }\href {\doibase
  10.1088/0022-3727/47/3/034009} {\bibfield  {journal} {\bibinfo  {journal} {J.
  Phys. D: Appl. Phys.}\ }\textbf {\bibinfo {volume} {47}} (\bibinfo {year}
  {2014}),\ 10.1088/0022-3727/47/3/034009}\BibitemShut {NoStop}%
\bibitem [{\citenamefont {Pandya}\ \emph {et~al.}(2016)\citenamefont {Pandya},
  \citenamefont {Damodaran}, \citenamefont {Xu}, \citenamefont {Hsu},
  \citenamefont {Agar},\ and\ \citenamefont {Martin}}]{pandya:16}%
  \BibitemOpen
  \bibfield  {author} {\bibinfo {author} {\bibfnamefont {S.}~\bibnamefont
  {Pandya}}, \bibinfo {author} {\bibfnamefont {A.~R.}\ \bibnamefont
  {Damodaran}}, \bibinfo {author} {\bibfnamefont {R.}~\bibnamefont {Xu}},
  \bibinfo {author} {\bibfnamefont {S.~L.}\ \bibnamefont {Hsu}}, \bibinfo
  {author} {\bibfnamefont {J.~C.}\ \bibnamefont {Agar}}, \ and\ \bibinfo
  {author} {\bibfnamefont {L.~W.}\ \bibnamefont {Martin}},\ }\href {\doibase
  10.1038/srep26075} {\bibfield  {journal} {\bibinfo  {journal} {Scientific
  Reports}\ }\textbf {\bibinfo {volume} {26075}} (\bibinfo {year} {2016}),\
  10.1038/srep26075}\BibitemShut {NoStop}%
\bibitem [{\citenamefont {Subramanian}\ \emph {et~al.}(1994)\citenamefont
  {Subramanian}, \citenamefont {Crawford},\ and\ \citenamefont
  {Harlow}}]{subramanian:94}%
  \BibitemOpen
  \bibfield  {author} {\bibinfo {author} {\bibfnamefont {M.~A.}\ \bibnamefont
  {Subramanian}}, \bibinfo {author} {\bibfnamefont {M.~K.}\ \bibnamefont
  {Crawford}}, \ and\ \bibinfo {author} {\bibfnamefont {R.~L.}\ \bibnamefont
  {Harlow}},\ }\href {\doibase 10.1016/0025-5408(94)90120-1} {\bibfield
  {journal} {\bibinfo  {journal} {Materials Research Bulletin}\ }\textbf
  {\bibinfo {volume} {29}} (\bibinfo {year} {1994}),\
  10.1016/0025-5408(94)90120-1}\BibitemShut {NoStop}%
\bibitem [{\citenamefont {Groenendijk}\ \emph {et~al.}(2016)\citenamefont
  {Groenendijk}, \citenamefont {Manca}, \citenamefont {Mattoni}, \citenamefont
  {Kootstra}, \citenamefont {Gariglio}, \citenamefont {Huang}, \citenamefont
  {van Heumen},\ and\ \citenamefont {Caviglia}}]{groenendijk:16}%
  \BibitemOpen
  \bibfield  {author} {\bibinfo {author} {\bibfnamefont {D.~J.}\ \bibnamefont
  {Groenendijk}}, \bibinfo {author} {\bibfnamefont {N.}~\bibnamefont {Manca}},
  \bibinfo {author} {\bibfnamefont {G.}~\bibnamefont {Mattoni}}, \bibinfo
  {author} {\bibfnamefont {L.}~\bibnamefont {Kootstra}}, \bibinfo {author}
  {\bibfnamefont {S.}~\bibnamefont {Gariglio}}, \bibinfo {author}
  {\bibfnamefont {Y.}~\bibnamefont {Huang}}, \bibinfo {author} {\bibfnamefont
  {E.}~\bibnamefont {van Heumen}}, \ and\ \bibinfo {author} {\bibfnamefont
  {A.~D.}\ \bibnamefont {Caviglia}},\ }\href {\doibase 10.1063/1.4960101}
  {\bibfield  {journal} {\bibinfo  {journal} {Applied Physics Letters}\
  }\textbf {\bibinfo {volume} {109}} (\bibinfo {year} {2016}),\
  10.1063/1.4960101}\BibitemShut {NoStop}%
\bibitem [{\citenamefont {Nie}\ \emph {et~al.}(2015)\citenamefont {Nie},
  \citenamefont {King}, \citenamefont {Kim}, \citenamefont {Uchida},
  \citenamefont {Wei}, \citenamefont {Faeth}, \citenamefont {Ruff},
  \citenamefont {Xie}, \citenamefont {Pan}, \citenamefont {Fennie},
  \citenamefont {Schlom},\ and\ \citenamefont {Shen}}]{nie:15}%
  \BibitemOpen
  \bibfield  {author} {\bibinfo {author} {\bibfnamefont {Y.~F.}\ \bibnamefont
  {Nie}}, \bibinfo {author} {\bibfnamefont {P.~D.~C.}\ \bibnamefont {King}},
  \bibinfo {author} {\bibfnamefont {C.~H.}\ \bibnamefont {Kim}}, \bibinfo
  {author} {\bibfnamefont {M.}~\bibnamefont {Uchida}}, \bibinfo {author}
  {\bibfnamefont {H.~I.}\ \bibnamefont {Wei}}, \bibinfo {author} {\bibfnamefont
  {B.~D.}\ \bibnamefont {Faeth}}, \bibinfo {author} {\bibfnamefont {J.~P.}\
  \bibnamefont {Ruff}}, \bibinfo {author} {\bibfnamefont {L.}~\bibnamefont
  {Xie}}, \bibinfo {author} {\bibfnamefont {X.}~\bibnamefont {Pan}}, \bibinfo
  {author} {\bibfnamefont {C.}~\bibnamefont {Fennie}}, \bibinfo {author}
  {\bibfnamefont {D.~G.}\ \bibnamefont {Schlom}}, \ and\ \bibinfo {author}
  {\bibfnamefont {K.~M.}\ \bibnamefont {Shen}},\ }\href {\doibase
  10.1103/PhysRevLett.114.016401} {\bibfield  {journal} {\bibinfo  {journal}
  {Physical Review Letters}\ }\textbf {\bibinfo {volume} {114}} (\bibinfo
  {year} {2015}),\ 10.1103/PhysRevLett.114.016401}\BibitemShut {NoStop}%
\bibitem [{\citenamefont {Greaves}\ \emph {et~al.}(2011)\citenamefont
  {Greaves}, \citenamefont {Greer}, \citenamefont {Lakes},\ and\ \citenamefont
  {Rouxel}}]{greaves:11}%
  \BibitemOpen
  \bibfield  {author} {\bibinfo {author} {\bibfnamefont {G.~N.}\ \bibnamefont
  {Greaves}}, \bibinfo {author} {\bibfnamefont {A.~L.}\ \bibnamefont {Greer}},
  \bibinfo {author} {\bibfnamefont {R.~S.}\ \bibnamefont {Lakes}}, \ and\
  \bibinfo {author} {\bibfnamefont {T.}~\bibnamefont {Rouxel}},\ }\href
  {\doibase 10.1038/NMAT3134} {\bibfield  {journal} {\bibinfo  {journal}
  {Nature Materials}\ }\textbf {\bibinfo {volume} {10}} (\bibinfo {year}
  {2011}),\ 10.1038/NMAT3134}\BibitemShut {NoStop}%
\bibitem [{\citenamefont {Huang}\ and\ \citenamefont
  {Chen}(2016)}]{huang_Poisson:16}%
  \BibitemOpen
  \bibfield  {author} {\bibinfo {author} {\bibfnamefont {C.}~\bibnamefont
  {Huang}}\ and\ \bibinfo {author} {\bibfnamefont {L.}~\bibnamefont {Chen}},\
  }\href {\doibase 10.1002/adma.201601363} {\bibfield  {journal} {\bibinfo
  {journal} {Advanced Materials}\ }\textbf {\bibinfo {volume} {28}} (\bibinfo
  {year} {2016}),\ 10.1002/adma.201601363}\BibitemShut {NoStop}%
\bibitem [{\citenamefont {Iglesias}\ \emph {et~al.}(2017)\citenamefont
  {Iglesias}, \citenamefont {Sarantopoulos}, \citenamefont {Magen},\ and\
  \citenamefont {Rivadulla}}]{iglesias:17}%
  \BibitemOpen
  \bibfield  {author} {\bibinfo {author} {\bibfnamefont {L.}~\bibnamefont
  {Iglesias}}, \bibinfo {author} {\bibfnamefont {A.}~\bibnamefont
  {Sarantopoulos}}, \bibinfo {author} {\bibfnamefont {C.}~\bibnamefont
  {Magen}}, \ and\ \bibinfo {author} {\bibfnamefont {F.}~\bibnamefont
  {Rivadulla}},\ }\href {\doibase 10.1103/PhysRevB.95.165138} {\bibfield
  {journal} {\bibinfo  {journal} {Physical Review B}\ }\textbf {\bibinfo
  {volume} {95}} (\bibinfo {year} {2017}),\
  10.1103/PhysRevB.95.165138}\BibitemShut {NoStop}%
\bibitem [{\citenamefont {Aschauer}\ \emph {et~al.}(2013)\citenamefont
  {Aschauer}, \citenamefont {Pfenninger}, \citenamefont {Selbach},
  \citenamefont {Grande},\ and\ \citenamefont {Spaldin}}]{aschauer:13}%
  \BibitemOpen
  \bibfield  {author} {\bibinfo {author} {\bibfnamefont {U.}~\bibnamefont
  {Aschauer}}, \bibinfo {author} {\bibfnamefont {R.}~\bibnamefont
  {Pfenninger}}, \bibinfo {author} {\bibfnamefont {S.~M.}\ \bibnamefont
  {Selbach}}, \bibinfo {author} {\bibfnamefont {T.}~\bibnamefont {Grande}}, \
  and\ \bibinfo {author} {\bibfnamefont {N.~A.}\ \bibnamefont {Spaldin}},\
  }\href {\doibase 10.1103/PhysRevB.88.054111} {\bibfield  {journal} {\bibinfo
  {journal} {Physical Review B}\ }\textbf {\bibinfo {volume} {88}} (\bibinfo
  {year} {2013}),\ 10.1103/PhysRevB.88.054111}\BibitemShut {NoStop}%
\bibitem [{\citenamefont {Chikara}\ \emph {et~al.}(2010)\citenamefont
  {Chikara}, \citenamefont {Korneta}, \citenamefont {Crummett}, \citenamefont
  {DeLong}, \citenamefont {Schlottmann},\ and\ \citenamefont
  {Cao}}]{chikara:10}%
  \BibitemOpen
  \bibfield  {author} {\bibinfo {author} {\bibfnamefont {S.}~\bibnamefont
  {Chikara}}, \bibinfo {author} {\bibfnamefont {O.}~\bibnamefont {Korneta}},
  \bibinfo {author} {\bibfnamefont {W.~P.}\ \bibnamefont {Crummett}}, \bibinfo
  {author} {\bibfnamefont {L.~E.}\ \bibnamefont {DeLong}}, \bibinfo {author}
  {\bibfnamefont {P.}~\bibnamefont {Schlottmann}}, \ and\ \bibinfo {author}
  {\bibfnamefont {G.}~\bibnamefont {Cao}},\ }\href {\doibase 10.1063/1.3362912}
  {\bibfield  {journal} {\bibinfo  {journal} {Journal of Applied Physics}\
  }\textbf {\bibinfo {volume} {107}} (\bibinfo {year} {2010}),\
  10.1063/1.3362912}\BibitemShut {NoStop}%
\bibitem [{\citenamefont {Shimura}\ \emph {et~al.}(1995)\citenamefont
  {Shimura}, \citenamefont {Inaguma}, \citenamefont {Nakamura}, \citenamefont
  {Itoh},\ and\ \citenamefont {Morii}}]{shimura:95}%
  \BibitemOpen
  \bibfield  {author} {\bibinfo {author} {\bibfnamefont {T.}~\bibnamefont
  {Shimura}}, \bibinfo {author} {\bibfnamefont {Y.}~\bibnamefont {Inaguma}},
  \bibinfo {author} {\bibfnamefont {T.}~\bibnamefont {Nakamura}}, \bibinfo
  {author} {\bibfnamefont {M.}~\bibnamefont {Itoh}}, \ and\ \bibinfo {author}
  {\bibfnamefont {Y.}~\bibnamefont {Morii}},\ }\href {\doibase
  10.1103/PhysRevB.52.9143} {\bibfield  {journal} {\bibinfo  {journal}
  {Physical Review B}\ }\textbf {\bibinfo {volume} {52}} (\bibinfo {year}
  {1995}),\ 10.1103/PhysRevB.52.9143}\BibitemShut {NoStop}%
\bibitem [{\citenamefont {Boseggia}\ \emph {et~al.}(2013)\citenamefont
  {Boseggia}, \citenamefont {Walker}, \citenamefont {Vale}, \citenamefont
  {Springell}, \citenamefont {Feng}, \citenamefont {Perry}, \citenamefont
  {Sala}, \citenamefont {H.~M.~Ronnow},\ and\ \citenamefont
  {McMorrow}}]{boseggia:13}%
  \BibitemOpen
  \bibfield  {author} {\bibinfo {author} {\bibfnamefont {S.}~\bibnamefont
  {Boseggia}}, \bibinfo {author} {\bibfnamefont {H.~C.}\ \bibnamefont
  {Walker}}, \bibinfo {author} {\bibfnamefont {J.}~\bibnamefont {Vale}},
  \bibinfo {author} {\bibfnamefont {R.}~\bibnamefont {Springell}}, \bibinfo
  {author} {\bibfnamefont {Z.}~\bibnamefont {Feng}}, \bibinfo {author}
  {\bibfnamefont {R.~S.}\ \bibnamefont {Perry}}, \bibinfo {author}
  {\bibfnamefont {M.~M.}\ \bibnamefont {Sala}}, \bibinfo {author}
  {\bibfnamefont {S.~P.~C.}\ \bibnamefont {H.~M.~Ronnow}}, \ and\ \bibinfo
  {author} {\bibfnamefont {D.~F.}\ \bibnamefont {McMorrow}},\ }\href {\doibase
  10.1088/0953-8984/25/42/422202} {\bibfield  {journal} {\bibinfo  {journal}
  {Journal of Physics: Condensed Matter}\ }\textbf {\bibinfo {volume} {25}}
  (\bibinfo {year} {2013}),\ 10.1088/0953-8984/25/42/422202}\BibitemShut
  {NoStop}%
\bibitem [{\citenamefont {Ge}\ \emph {et~al.}(2011)\citenamefont {Ge},
  \citenamefont {Qi}, \citenamefont {Korneta}, \citenamefont {Long},
  \citenamefont {Schlottmann}, \citenamefont {Crummett}, ,\ and\ \citenamefont
  {Cao}}]{ge:11}%
  \BibitemOpen
  \bibfield  {author} {\bibinfo {author} {\bibfnamefont {M.}~\bibnamefont
  {Ge}}, \bibinfo {author} {\bibfnamefont {T.~F.}\ \bibnamefont {Qi}}, \bibinfo
  {author} {\bibfnamefont {O.~B.}\ \bibnamefont {Korneta}}, \bibinfo {author}
  {\bibfnamefont {D.~E.~D.}\ \bibnamefont {Long}}, \bibinfo {author}
  {\bibfnamefont {P.}~\bibnamefont {Schlottmann}}, \bibinfo {author}
  {\bibfnamefont {W.~P.}\ \bibnamefont {Crummett}}, , \ and\ \bibinfo {author}
  {\bibfnamefont {G.}~\bibnamefont {Cao}},\ }\href {\doibase
  10.1103/PhysRevB.84.100402} {\bibfield  {journal} {\bibinfo  {journal}
  {Physical Review B}\ }\textbf {\bibinfo {volume} {84}} (\bibinfo {year}
  {2011}),\ 10.1103/PhysRevB.84.100402}\BibitemShut {NoStop}%
\bibitem [{\citenamefont {Ye}\ \emph {et~al.}(2013)\citenamefont {Ye},
  \citenamefont {Chi}, \citenamefont {Chakoumakos}, \citenamefont
  {Fernandez-Baca}, \citenamefont {Qi},\ and\ \citenamefont {Cao}}]{ye:13}%
  \BibitemOpen
  \bibfield  {author} {\bibinfo {author} {\bibfnamefont {F.}~\bibnamefont
  {Ye}}, \bibinfo {author} {\bibfnamefont {S.}~\bibnamefont {Chi}}, \bibinfo
  {author} {\bibfnamefont {B.~C.}\ \bibnamefont {Chakoumakos}}, \bibinfo
  {author} {\bibfnamefont {J.~A.}\ \bibnamefont {Fernandez-Baca}}, \bibinfo
  {author} {\bibfnamefont {T.}~\bibnamefont {Qi}}, \ and\ \bibinfo {author}
  {\bibfnamefont {G.}~\bibnamefont {Cao}},\ }\href {\doibase
  10.1103/PhysRevB.87.140406} {\bibfield  {journal} {\bibinfo  {journal}
  {Physical Review B}\ }\textbf {\bibinfo {volume} {87}} (\bibinfo {year}
  {2013}),\ 10.1103/PhysRevB.87.140406}\BibitemShut {NoStop}%
\bibitem [{\citenamefont {Miao}\ \emph {et~al.}(2014)\citenamefont {Miao},
  \citenamefont {Xu},\ and\ \citenamefont {Mao}}]{miao:14}%
  \BibitemOpen
  \bibfield  {author} {\bibinfo {author} {\bibfnamefont {L.}~\bibnamefont
  {Miao}}, \bibinfo {author} {\bibfnamefont {H.}~\bibnamefont {Xu}}, \ and\
  \bibinfo {author} {\bibfnamefont {Z.~Q.}\ \bibnamefont {Mao}},\ }\href
  {\doibase 10.1103/PhysRevB.89.035109} {\bibfield  {journal} {\bibinfo
  {journal} {Physical Review B}\ }\textbf {\bibinfo {volume} {89}} (\bibinfo
  {year} {2014}),\ 10.1103/PhysRevB.89.035109}\BibitemShut {NoStop}%
\bibitem [{\citenamefont {Cao}\ \emph {et~al.}(2002)\citenamefont {Cao},
  \citenamefont {Xin}, \citenamefont {Alexander}, \citenamefont {Crow},
  \citenamefont {Schlottmann}, \citenamefont {Crawford}, \citenamefont
  {Harlow},\ and\ \citenamefont {Marshall}}]{cao:02}%
  \BibitemOpen
  \bibfield  {author} {\bibinfo {author} {\bibfnamefont {G.}~\bibnamefont
  {Cao}}, \bibinfo {author} {\bibfnamefont {Y.}~\bibnamefont {Xin}}, \bibinfo
  {author} {\bibfnamefont {C.~S.}\ \bibnamefont {Alexander}}, \bibinfo {author}
  {\bibfnamefont {J.~E.}\ \bibnamefont {Crow}}, \bibinfo {author}
  {\bibfnamefont {P.}~\bibnamefont {Schlottmann}}, \bibinfo {author}
  {\bibfnamefont {M.~K.}\ \bibnamefont {Crawford}}, \bibinfo {author}
  {\bibfnamefont {R.}~\bibnamefont {Harlow}}, \ and\ \bibinfo {author}
  {\bibfnamefont {W.}~\bibnamefont {Marshall}},\ }\href {\doibase
  10.1103/PhysRevB.66.214412} {\bibfield  {journal} {\bibinfo  {journal}
  {Physical Review B}\ }\textbf {\bibinfo {volume} {66}} (\bibinfo {year}
  {2002}),\ 10.1103/PhysRevB.66.214412}\BibitemShut {NoStop}%
\bibitem [{\citenamefont {Kini}\ \emph {et~al.}(2006)\citenamefont {Kini},
  \citenamefont {Strydom}, \citenamefont {Jeevan}, \citenamefont {Geibel}, ,\
  and\ \citenamefont {Ramakrishnan}}]{kini:06}%
  \BibitemOpen
  \bibfield  {author} {\bibinfo {author} {\bibfnamefont {N.~S.}\ \bibnamefont
  {Kini}}, \bibinfo {author} {\bibfnamefont {A.~M.}\ \bibnamefont {Strydom}},
  \bibinfo {author} {\bibfnamefont {H.~S.}\ \bibnamefont {Jeevan}}, \bibinfo
  {author} {\bibfnamefont {C.}~\bibnamefont {Geibel}}, , \ and\ \bibinfo
  {author} {\bibfnamefont {S.}~\bibnamefont {Ramakrishnan}},\ }\href {\doibase
  10.1088/0953-8984/18/35/008} {\bibfield  {journal} {\bibinfo  {journal}
  {Journal of Physics: Condensed Matter}\ }\textbf {\bibinfo {volume} {18}},\
  \bibinfo {pages} {8205} (\bibinfo {year} {2006})}\BibitemShut {NoStop}%
\bibitem [{\citenamefont {Biswas}\ \emph {et~al.}(2014)\citenamefont {Biswas},
  \citenamefont {Kim},\ and\ \citenamefont {Jeong}}]{biswas:14}%
  \BibitemOpen
  \bibfield  {author} {\bibinfo {author} {\bibfnamefont {A.}~\bibnamefont
  {Biswas}}, \bibinfo {author} {\bibfnamefont {K.~S.}\ \bibnamefont {Kim}}, \
  and\ \bibinfo {author} {\bibfnamefont {Y.~H.}\ \bibnamefont {Jeong}},\ }\href
  {\doibase 10.1063/1.4903314} {\bibfield  {journal} {\bibinfo  {journal}
  {Journal of Applied Physics}\ }\textbf {\bibinfo {volume} {116}} (\bibinfo
  {year} {2014}),\ 10.1063/1.4903314}\BibitemShut {NoStop}%
\bibitem [{\citenamefont {Matsuno}\ \emph {et~al.}(2015)\citenamefont
  {Matsuno}, \citenamefont {Ihara}, \citenamefont {Yamamura}, \citenamefont
  {Wadati}, \citenamefont {Ishii}, \citenamefont {Shankar}, \citenamefont
  {Kee},\ and\ \citenamefont {Takagi}}]{matsuno:15}%
  \BibitemOpen
  \bibfield  {author} {\bibinfo {author} {\bibfnamefont {J.}~\bibnamefont
  {Matsuno}}, \bibinfo {author} {\bibfnamefont {K.}~\bibnamefont {Ihara}},
  \bibinfo {author} {\bibfnamefont {S.}~\bibnamefont {Yamamura}}, \bibinfo
  {author} {\bibfnamefont {H.}~\bibnamefont {Wadati}}, \bibinfo {author}
  {\bibfnamefont {K.}~\bibnamefont {Ishii}}, \bibinfo {author} {\bibfnamefont
  {V.~V.}\ \bibnamefont {Shankar}}, \bibinfo {author} {\bibfnamefont {H.~Y.}\
  \bibnamefont {Kee}}, \ and\ \bibinfo {author} {\bibfnamefont
  {H.}~\bibnamefont {Takagi}},\ }\href {\doibase
  10.1103/PhysRevLett.114.247209} {\bibfield  {journal} {\bibinfo  {journal}
  {Physical Review Letters}\ }\textbf {\bibinfo {volume} {114}} (\bibinfo
  {year} {2015}),\ 10.1103/PhysRevLett.114.247209}\BibitemShut {NoStop}%
\bibitem [{\citenamefont {Ohuchi}\ \emph {et~al.}(2018)\citenamefont {Ohuchi},
  \citenamefont {Matsuno}, \citenamefont {Ogawa}, \citenamefont {Kozuka},
  \citenamefont {Uchida}, \citenamefont {Tokura},\ and\ \citenamefont
  {Kawasaki}}]{ohuchi:18}%
  \BibitemOpen
  \bibfield  {author} {\bibinfo {author} {\bibfnamefont {Y.}~\bibnamefont
  {Ohuchi}}, \bibinfo {author} {\bibfnamefont {J.}~\bibnamefont {Matsuno}},
  \bibinfo {author} {\bibfnamefont {N.}~\bibnamefont {Ogawa}}, \bibinfo
  {author} {\bibfnamefont {Y.}~\bibnamefont {Kozuka}}, \bibinfo {author}
  {\bibfnamefont {M.}~\bibnamefont {Uchida}}, \bibinfo {author} {\bibfnamefont
  {Y.}~\bibnamefont {Tokura}}, \ and\ \bibinfo {author} {\bibfnamefont
  {M.}~\bibnamefont {Kawasaki}},\ }\href {\doibase 10.1038/s41467-017-02629-3}
  {\bibfield  {journal} {\bibinfo  {journal} {Nature Communications}\ }\textbf
  {\bibinfo {volume} {9}} (\bibinfo {year} {2018}),\
  10.1038/s41467-017-02629-3}\BibitemShut {NoStop}%
\bibitem [{\citenamefont {Tselev}\ \emph {et~al.}(2015)\citenamefont {Tselev},
  \citenamefont {Vasudevan}, \citenamefont {Gianfrancesco}, \citenamefont
  {Qiao}, \citenamefont {Ganesh}, \citenamefont {Meyer}, \citenamefont {Lee},
  \citenamefont {Biegalski}, \citenamefont {Baddorf},\ and\ \citenamefont
  {Kalinin}}]{tselev:15}%
  \BibitemOpen
  \bibfield  {author} {\bibinfo {author} {\bibfnamefont {A.}~\bibnamefont
  {Tselev}}, \bibinfo {author} {\bibfnamefont {R.~K.}\ \bibnamefont
  {Vasudevan}}, \bibinfo {author} {\bibfnamefont {A.~G.}\ \bibnamefont
  {Gianfrancesco}}, \bibinfo {author} {\bibfnamefont {L.}~\bibnamefont {Qiao}},
  \bibinfo {author} {\bibfnamefont {P.}~\bibnamefont {Ganesh}}, \bibinfo
  {author} {\bibfnamefont {T.~L.}\ \bibnamefont {Meyer}}, \bibinfo {author}
  {\bibfnamefont {H.~N.}\ \bibnamefont {Lee}}, \bibinfo {author} {\bibfnamefont
  {M.~D.}\ \bibnamefont {Biegalski}}, \bibinfo {author} {\bibfnamefont {A.~P.}\
  \bibnamefont {Baddorf}}, \ and\ \bibinfo {author} {\bibfnamefont {S.~V.}\
  \bibnamefont {Kalinin}},\ }\href {\doibase 10.1021/acsnano.5b00743}
  {\bibfield  {journal} {\bibinfo  {journal} {ACS Nano}\ }\textbf {\bibinfo
  {volume} {9}} (\bibinfo {year} {2015}),\ 10.1021/acsnano.5b00743}\BibitemShut
  {NoStop}%
\end{thebibliography}

%

\end{document}